\documentclass{aa}

\usepackage{amsmath}
\usepackage{slashed}
\usepackage{bbold}
\usepackage{natbib}
\usepackage{graphicx}
\usepackage{tabularx}
\usepackage{url}

\bibpunct{(}{)}{;}{a}{}{,}

\begin{document}
 
\title{Log-transforming the matter power spectrum}

\author{M.~Greiner\thanks{\email{maksim@mpa-garching.mpg.de}}
  \and T.A.~En\ss{}lin}

\institute{Max-Planck-Institut f\"ur Astrophysik, Karl-Schwarzschild-Str.~1, 85748 Garching, Germany \label{inst:MPA}
\and Ludwig-Maximilians-Universit\"at M\"unchen, Geschwister-Scholl-Platz~1, 80539 M\"unchen, Germany \label{inst:LMU}}

\date{Received DD MMM. YYYY / Accepted DD MMM. YYYY}

\abstract{
We investigate whether non-linear effects on the large-scale power spectrum of dark matter, namely the increase in small-scale power and the smearing of baryon acoustic oscillations, can be decreased by a log-transformation or emulated by an exponential transformation of the linear spectrum. To that end we present a formalism to convert the power spectrum of a log-normal field to the power spectrum of the logarithmic Gaussian field and vice versa. All ingredients of our derivation can already be found in various publications in cosmology and other fields. We follow a more pedagogical approach providing a detailed derivation, application examples, and a discussion of implementation subtleties in one text. We use the formalism to show that the non-linear increase in small-scale power in the matter power spectrum is significantly smaller for the log-transformed spectrum which fits the linear spectrum (with less than 20\% error) for redshifts down to 1 and $k\leq1.0\,h\,\mathrm{Mpc}$. For lower redshifts the fit to the linear spectrum is not as good, but the reduction of non-linear effects is still significant. Similarly, we show that applying the linear growth factor to the logarithmic density leads to an automatic increase in small-scale power for low redshifts fitting to third-order perturbation spectra and Cosmic Emulator spectra with an error of less than $20\%$. Smearing of baryon acoustic oscillations is at least three times weaker, but still present.
}


\titlerunning{Log-transforming the matter power spectrum}
\authorrunning{M.~Greiner and T.A.~En\ss{}lin}
\maketitle

\section{Introduction}

In cosmology and astrophysics densities are often described as a combination of the mean density and fluctuations around it,
\begin{equation}
 \rho = \rho_0\left( 1 + \delta \right),
 \label{eq:contrast}
\end{equation}
where the density contrast field $\delta$ is often small and accurately described by a Gaussian random field. However, as soon as large overdensities occur ($\delta>1$) a description of $\delta$ by Gaussian statistics does not suffice, since $\delta$ cannot go below minus one.
A prominent example of such a density is the large-scale structure, where the overdensity is accurately described by Gaussian statistics at early epochs, but becomes highly non-linear at later epochs and small scales. In these non-linear regimes the density is much more accurately described by log-normal statistics, i.e., $\log(1+\delta)$ following Gaussian statistics. This has already been noted by \cite{Hubble-1934} and \cite{Hamilton-1985} and theoretically investigated by \cite{Coles-1991}. N-body simulations calculated by \cite{Kayo-2001} are also in agreement with a log-normal distribution. However, as \cite{Carron-2011} showed, correlation functions and spectra contain very limited information in the case of highly non-linear log-normal statistics as there are other distribution functions that would produce the same moments. For a perturbative treatment of the power spectrum of $\log(1+\delta)$ see \cite{Wang-2011}.
Another example is the density in turbulent and isothermal clouds, which is very accurately described by a log-normal distribution (seem e.g., \cite{Passot-1998} or \cite{Nordlund-1999}).

In many cases, however, only the linear power spectrum (the power spectrum of $\delta$) is available (e.g., from observations or simulations). It is therefore useful to convert a linear power spectrum into a logarithmic power spectrum (and vice versa). In this work we derive such a conversion formula under the assumption that the power spectrum is the power spectrum of the underlying statistical process. This condition is not necessarily met if the power spectrum is calculated from observational data or simulations, but in most cases the formula works nevertheless.
Formulas relating the correlation functions of Gaussian and log-normal fields are widely known throughout several fields. In cosmology we refer the reader to \cite{Coles-1991}, \cite{Politzer-1984}, and \cite{Percival-2004}. The first goal of this paper is to provide a pedagogical introduction along with application examples enabling the reader to directly apply the derived formulas in a numerical setting while avoiding normalization and prefactor mistakes which can drastically influence the outcome due to the non-linearity of the exponential and logarithmic functions. The second goal is to apply the presented formalism to investigate how much a log-transformation of the large-scale density reduces the non-linear enhancement of small-scale power for decreasing redshifts.

The remainder of this paper is structured as follows. First, we go through the derivation of the conversion formalism in Sect.~\ref{sec:conversion} in which we state all definitions explicitly before the calculation. Second, we apply the conversion to the large-scale matter spectrum in Sect.~\ref{sec:matter_spectrum} to test the validity of the log-normal approximation of the cosmic matter density. Finally, we conclude this paper with a summary and outlook in Sect.~\ref{sec:conclusions}.

In the Appendix, we describe how to generate log-normal random field following a given spectrum (Appendix~\ref{sec:generate_lognormal}). Furthermore, we present a consistent discretization of the conversion formulas (Appendix~\ref{sec:discrete}) their spherical harmonics equivalent (Appendix~\ref{sec:spherical}). We also discuss in greater detail a number of aspects from Sect.~\ref{sec:matter_spectrum} in Appendix.~\ref{sec:matter_appendix}, e.g., how the choice of gridding affects the results.

\section{The conversion}
\label{sec:conversion}

\subsection{Notation and definitions}
\label{sec:definitions}

We denote the log-normal field at position $\vec{x}$ by $\rho(\vec{x})$ and the underlying Gaussian field by $s(\vec{x})$,
\begin{equation}
\begin{split}
\rho(\vec{x}) = \rho_0\, e^{s(\vec{x})} = e^{s(\vec{x})+m},
\end{split}
\end{equation}
where $s$ is dimensionless and the units of $\rho$ as well as a proportionality constant are in $\rho_0$ and $m \equiv \log \rho_0$. In this work, $\log$ denotes the natural logarithm.

The field $s$ is assumed to follow a Gaussian distribution with zero mean\footnote{A non-zero mean can be absorbed into $m$.},
\begin{equation}
\mathcal{P}(s) =  \frac{1}{Z} \exp\!\left(-\frac{1}{2}\int\!\!\mathrm{d}^ux\,\mathrm{d}^uy\ s(\vec{x}) \left(S^{-1}\right)\!\!(\vec{x},\vec{y})\, s(\vec{y})  \right),
\end{equation}
with $u$ being the dimensionality of space and $Z = \mathrm{det}(2 \pi S)^{\frac{1}{2}}$. $S(\vec{x},\vec{y})$ is the auto-correlation function of $s$. It is defined as
\begin{equation}
\begin{split}
 S(\vec{x},\vec{y}) & :=  \left\langle s(\vec{x}) s(\vec{y}) \right\rangle_{\mathcal{P}(s)} \equiv \int\!\!\mathcal{D}s\ \mathcal{P}(s)\, s(\vec{x}) s(\vec{y}).
\end{split}
\end{equation}
Here, $\mathcal{D}s$ denotes integration over the full phase space of $s$. The inverse of $S$ is defined by the relation
\begin{equation}
 \int\!\!\mathrm{d}^uy\ S(\vec{x},\vec{y})\, \left(S^{-1}\right)\!\!(\vec{y},\vec{z}) = \delta_\mathrm{D}(\vec{x}-\vec{z}),
\end{equation}
where $\delta_\mathrm{D}(\vec{x}-\vec{z})$ is the $u$-dimensional Dirac delta distribution.

In Fourier space, we denote position vectors by $\vec{k}$ or $\vec{q}$. Fourier transformed fields are denoted by their argument (e.g.,~$s(\vec{k})$).
We define the Fourier transformations over continuous position space as
\begin{equation}
 \begin{split}
  s(\vec{k}) = \int\!\!\mathrm{d}^ux\ e^{i \vec{k}\cdot\vec{x}}\, s(\vec{x}),\quad
  s(\vec{x}) = \int\!\!\frac{\mathrm{d}^uk}{(2\pi)^u}\ e^{-i \vec{k}\cdot\vec{x}}\, s(\vec{k}).
 \end{split}
\end{equation}

If a field obeys statistical homogeneity its auto-correlation function depends only on the separation, $S(\vec{x},\vec{y}) = S(\vec{x}-\vec{y})$,
and is therefore diagonal in Fourier space,
\begin{equation}
\begin{split}
 S(\vec{k},\vec{q}) & = \int\!\!\mathrm{d}^ux\,\mathrm{d}^uy\ e^{i\vec{k}\cdot\vec{x} - i\vec{q}\cdot\vec{y}}\ \left\langle s(\vec{x}) s(\vec{y}) \right\rangle_{\mathcal{P}(s)}\\
& = (2\pi)^u \delta_\mathrm{D}(\vec{k}-\vec{q})\, P_s(\vec{k}).
\end{split}
\label{eq:first_powspec}
\end{equation}
The quantity $P_s(\vec{k})$ is the statistical power spectrum of $s$.

The auto-correlation function and statistical power spectrum of $\rho$ are defined analogously,
\begin{equation}
\begin{split}
 C_\rho(\vec{x},\vec{y}) & = \left\langle \rho(\vec{x}) \rho(\vec{y}) \right\rangle_{\mathcal{P}(\rho)} = \rho_0^2\left\langle e^{s(\vec{x})} e^{s(\vec{y})} \right\rangle_{\mathcal{P}(s)},\\
 C_\rho(\vec{k},\vec{q}) & = (2\pi)^u \delta_\mathrm{D}(\vec{k}-\vec{q})\, P_\rho(\vec{k}).
\label{eq:second_powspec}
\end{split}
\end{equation}
In the second line we used that $\rho$ obeys statistical homogeneity, too.

\subsection{Converting the logarithmic power spectrum to the linear power spectrum -- the forward conversion}
\label{sec:forward}

Suppose the statistical power spectrum of $s$, $P_s(\vec{k})$, and the mean $m$ are known. We are looking for the statistical power spectrum of \mbox{$\rho(\vec{x}) = e^{s(\vec{x})+m}$}. The auto-correlation functions are related by
\begin{equation}
\begin{split}
C_\rho(\vec{x},\vec{y}) & = \left\langle e^{s(\vec{x})+m}\, e^{s(\vec{y})+m} \right\rangle_{\mathcal{P}(s)}\\
& = e^{2m + \frac{1}{2}S(\vec{x},\vec{x}) + \frac{1}{2}S(\vec{y},\vec{y}) + S(\vec{x},\vec{y})},
\end{split}
\end{equation}
since $\mathcal{P}(s)$ is Gaussian. Using the statistical homogeneity of $C_\rho$ and $S$, we write
\begin{equation}
 C_\rho(\vec{0},\vec{x}) = e^{2m + S(\vec{0},\vec{0}) + S(\vec{0},\vec{x})}.
\end{equation}
This relation between the correlation function of a Gaussian and log-normal field is well known in cosmology (see, e.g., \cite{Coles-1991}).
Combined with Eqs.~\eqref{eq:first_powspec} and \eqref{eq:second_powspec} this yields
\begin{equation}
\begin{split}
&\int\!\!\frac{\mathrm{d}^uk}{(2\pi)^u}\ e^{-i \vec{k}\cdot\vec{x}} \ P_\rho(\vec{k})\\ & = \exp\!\left(2m + \int\!\!\frac{\mathrm{d}^uk}{(2\pi)^u} \left(e^{-i \vec{k}\cdot\vec{x}} + 1\right) P_s(\vec{k})  \right),
\end{split}
\label{eq:relation_between_spectra}
\end{equation}
and therefore
\begin{equation}
\begin{split}
 P_\rho(\vec{k}) = & \int\!\!\mathrm{d}^ux\ e^{i \vec{k}\cdot\vec{x}}\ e^{2m} \\
 & \times \exp\!\left(\int\!\!\frac{\mathrm{d}^uq}{(2\pi)^u} \left(e^{-i \vec{q}\cdot\vec{x}} + 1\right) P_s(\vec{q}) \right).
\end{split}
\label{eq:forward_conv}
\end{equation}
Eq.~\eqref{eq:forward_conv} is our \textit{forward conversion} formula.

If $P_s(\vec{k})$ is isotropic, the equation can be simplified in spherical coordinates, since in that case the argument of the exponential function is isotropic in $\vec{x}$, too. In the three-dimensional isotropic case, integration of the angular part yields
\begin{equation}
\begin{split}
 P_\rho(k) =\ & 4\pi\int\limits_0^\infty\!\!\mathrm{d}r\ r^2 \frac{\sin(kr)}{kr} e^{2m}\\
& \times  \exp\!\left( \int\limits_0^\infty\!\!\frac{\mathrm{d}q}{2\pi^2} q^2 \left(\frac{\sin(qr)}{qr} + 1\right) P_s(q)\right),
\end{split}
\label{eq:angles_integrated}
\end{equation}
where we denoted $P(k)\equiv \left.P(\vec{q})\right|_{|\vec{q}|\!=\!k}$.
While this equation only has one-dimensional integrals, its numerical evaluation involves some subtleties, which is why we perform the numerical calculations in this work using full-dimensional fast Fourier transforms, even if the power spectrum is isotropic. However, integration of the angular part can be performed for all of the following equations in a very similar manner.

\subsection{Converting the linear power spectrum to the logarithmic power spectrum -- the backward conversion}
\label{sec:backward}

Suppose the statistical power spectrum of $\rho$, $P_\rho(k)$, is known and we want to know the power spectrum of $s$.
We therefore need to solve Eq.~\eqref{eq:relation_between_spectra} for $P_s(k)$,
\begin{equation}
\begin{split}
&\log\!\left(\int\!\!\frac{\mathrm{d}^uk}{(2\pi)^u}\ e^{-i \vec{k}\cdot\vec{x}} \ P_\rho(\vec{k})\right) -2m\\
& = \int\!\!\frac{\mathrm{d}^uk}{(2\pi)^u} \left(e^{-i \vec{k}\cdot\vec{x}} + 1\right) P_s(\vec{k}).
\end{split}
\end{equation}
We invert the operation in front of $P_s(k)$ using
\begin{equation}
\begin{split}
 & \int\!\!\mathrm{d}^ux \left( e^{i \vec{q}\cdot\vec{x}} - \frac{1}{2}\delta_\mathrm{D}(\vec{x})\,(2\pi)^u\delta_\mathrm{D}(\vec{q}) \right) \left( e^{-i \vec{k}\cdot\vec{x}} + 1 \right) \\
& = (2\pi)^u\delta_\mathrm{D}(\vec{k}-\vec{q}),
\end{split}
\end{equation}
to arrive at 
\begin{equation}
\begin{split}
  P_s(\vec{k}) =\ & \int\!\!\mathrm{d}^ux\ e^{i \vec{k}\cdot\vec{x}} \log\!\left( \int\!\!\frac{\mathrm{d}^uq}{(2\pi)^u}\ e^{-i \vec{q}\cdot\vec{x}} \ P_\rho(\vec{q}) \right)\\
 & - m\,(2\pi)^u\delta_\mathrm{D}(k)\\
 & - \frac{1}{2}\ (2\pi)^u\delta_\mathrm{D}(k)\, \log\!\left( \int\!\!\frac{\mathrm{d}^uq}{(2\pi)^u}\ P_\rho(\vec{q}) \right).
\end{split}
\end{equation}
The two correction terms on the right-hand-side do not affect modes with $\vec{k}\!\neq\!\vec{0}$,
\begin{equation}
 P_s(\vec{k}\!\neq\!\vec{0})  = \int\!\!\mathrm{d}^ux\ e^{i \vec{k}\cdot\vec{x}} \log\!\left( \int\!\!\frac{\mathrm{d}^uq}{(2\pi)^u}\ e^{-i \vec{q}\cdot\vec{x}} \ P_\rho(\vec{q}) \right),
\label{eq:backward_no_mono}
\end{equation}
and there is a degeneracy between the monopole $P_s(\vec{k}\!=\!\vec{0})$ and the mean,
\begin{equation}
\begin{split}
 P_s(\vec{k}\!=\!\vec{0}) + Vm =\ & \int\!\!\mathrm{d}^ux\ \log\!\left( \int\!\!\frac{\mathrm{d}^uq}{(2\pi)^u}\ e^{-i \vec{q}\cdot\vec{x}} \ P_\rho(\vec{q}) \right)\\
& - \frac{V}{2}\log\!\left( \int\!\!\frac{\mathrm{d}^uq}{(2\pi)^u}\ P_\rho(\vec{q}) \right),
\end{split}
\label{eq:backward_mono}
\end{equation}
where we have identified $V = (2\pi)^u\delta_\mathrm{D}(\vec{0})$ as the volume of the system.
We recommend to set the monopole to zero to fix the mean.
Eqs.~\eqref{eq:backward_no_mono}~\&~\eqref{eq:backward_mono} are our \textit{backward conversion} formulas.

We note that knowledge about the monopole of $\rho$ is crucial for this conversion to work. If the monopole is not supplied, one can estimate it from the mean of $\rho$ as
\begin{equation}
  P_\rho(\vec{k}\!=\!\vec{0}) = V\left\langle \rho \right\rangle^2.
\label{eq:fill_mono}
\end{equation}
Using the assumption of statistical homogeneity the backward conversion formula can also be applied in real space to get $S(\vec{x},\vec{y})$ from $C_\rho(\vec{x},\vec{y})$,
\begin{equation}
 S(\vec{x},\vec{0}) = \log\!\left(C_\rho(\vec{x},\vec{0})\right) - \frac{1}{2} \log\!\left(C_\rho(\vec{0},\vec{0})\right) - m.
 \label{eq:real_space_backward}
\end{equation}
This relation can (in a different form) also be found in \cite{Coles-1991} (Eq.~(30) therein).

\section{The matter power spectrum}
\label{sec:matter_spectrum}

In this section, we use Eqs.~\eqref{eq:forward_conv},~\eqref{eq:backward_no_mono},~and~\eqref{eq:backward_mono} to test the range of validity of a log-normal approximation to the power spectrum of dark matter. In order to apply the formulas on a computer one needs to discretize them correctly, since global prefactors are important in a non-linear transformation such as the exponential and logarithmic functions. We present a consistent discretization in Appendix~\ref{sec:discrete}.

The cosmic matter density $\rho$ is typically parametrized as the mean density $\rho_0$ and mass density contrast $\delta$ according to Eq~.\eqref{eq:contrast}.
The three-dimensional isotropic power spectrum of these fluctuations, $P_\delta(k)$, is usually referred to as the matter power spectrum of the large-scale structure.
At high redshifts the fluctuations are small, $\delta \ll 1$. With decreasing redshifts, the magnitude of the fluctuations increases.
In linear theory, each Fourier mode of the fluctuations is enhanced by the same linear growth factor $D(z)$, yielding a very simple relation between the matter power spectrum at different redshifts,
\begin{equation}
D^2(z')\, P_\delta(k,z) = D^2(z)\, P_\delta(k,z'),
\end{equation}
where $P_\delta(k,z)$ denotes the power spectrum of $\delta$ at redshift $z$. The functional form of the linear growth factor is slightly different for different cosmologies. We use the cosmological parameters determined by \cite{Planck-2013}. The prefactor of the power spectrum is determined by the $\sigma_8$ normalization\footnote{This is the normalization of the linear matter power spectrum. If for example the $\left.\sigma_8\right|_{z=1}$ value is $0.51$ for the linear spectrum, it would be $0.55$ for the 3PT spectrum.} which is the variance of $\delta$ convolved with a spherical top hat function with an $8\,\mathrm{Mpc}/h$ radius. Its value at redshift zero is determined by \cite{Planck-2013} to be $\left.\sigma_8\right|_{z=0} = 0.83$. Together with the growth factor this determines the linear power spectrum $P_\delta(k)$ at all redshifts, since in linear theory it retains its shape.
However, the linear description of the redshift dependence of the matter power spectrum fails for low redshifts, where the fluctuations can be on the order of 1. This has been successfully treated up to redshift 1 by third order perturbation theory (henceforth 3PT) around the linearly evolved spectrum by \cite{Jeong-2006}. For redshifts below 1, the non-linearities in the matter power spectrum can be modeled using the Cosmic emulator (henceforth CosmicEmu) based on \cite{Heitmann-2009}, \cite{Heitmann-2010} and \cite{Lawrence-2010}.

Using the formalism presented in this paper we investigate whether some of the non-linearities resolved by third order perturbation theory and the Cosmic emulator arise naturally if linear growth is applied to the logarithmic density contrast instead of the density contrast itself. A similar question has been investigated by \cite{Neyrinck-2009}. Using data from the Millennium Simulation by \cite{Springel-2005} \cite{Neyrinck-2009} compared the power spectrum of $\log\!\left( 1+\delta \right)$ at different redshifts. They found a remarkable reduction of non-linearities in the power spectra up to $k=1.0\,h\,\mathrm{Mpc}^{-1}$ at all redshifts. They also found a bias factor for large scales, which is not apparent in our figures. We discuss the origin of this factor in Appendix~\ref{sec:bias-factor}.

Around redshift 7 the 3PT corrections by \cite{Jeong-2006} start to become significant. \cite{Jeong-2006} find a good agreement between the 3PT spectrum and N-body simulations for wavevectors up to $k=1.4\, h\, \mathrm{Mpc}^{-1}$ and redshifts higher or equal to 1. For redshifts $1$ to $0$ we use matter spectra calculated using CosmicEmu. According to \cite{Heitmann-2009}, \cite{Heitmann-2010} and \cite{Lawrence-2010} these are accurate to $1\%$ up to $k=1.0\, h\, \mathrm{Mpc}^{-1}$. We therefore assume, that they are also reasonably accurate up to $k=1.4\, h\, \mathrm{Mpc}^{-1}$.

We denote the logarithmic density contrast as
\begin{equation}
 s = \log\!\left( 1+\delta \right).
\end{equation}
At high redshifts, where $\delta \ll 1$, we have $s \approx \delta$. Therefore, at high redshifts applying the linear growth factor to $s$ has the same effect as applying it to $\delta$.
We keep applying the growth factor to $s$ instead of $\delta$ throughout the whole redshift spectrum in order to test whether some of the non-linearities appearing at lower redshifts arise naturally this way.
Therefore we apply the forward conversion formula (Eq.~\eqref{eq:forward_conv}) to the linearly evolved spectrum at redshifts $0$ to $7$ and the backward conversion formulas (Eqs.~\eqref{eq:backward_no_mono}~and~\eqref{eq:backward_mono}) to the 3PT spectra and the CosmicEmu spectra respectively.

To that end, we calculate the matter power spectrum at several redshifts between 1 and 7 by applying the 3PT code\footnote{\url{http://www.mpa-garching.mpg.de/~komatsu/CRL/powerspectrum/density3pt/}} by \cite{Jeong-2006} to the linear power spectrum calculated using CAMB\footnote{\url{http://camb.info/}} (see, e.g.,~\cite{Lewis-2000}) and between 0 and 1 using the CosmicEmu code\footnote{\url{http://www.lanl.gov/projects/cosmology/CosmicEmu/emu.html}}.

The lowest spectral length covered is $k = 0.004\,h\,\mathrm{Mpc}^{-1}$ for the 3PT code and $k = 0.0075\,h\,\mathrm{Mpc}^{-1}$ for the CosmicEmu code. Therefore, we let our numerical setup cover the region $0.0075\,h\,\mathrm{Mpc}^{-1} \leq k \leq 1.4\, h\, \mathrm{Mpc}^{-1}$.
There are several ways to resolve this spectral range and because of the non-linear nature of the exponential function the result of the conversion formulas is not independent of the choice. The smoothing onto a grid of a logarithmic function is thoroughly discussed by \cite{Wang-2011}. We perform the calculation on four different grids to demonstrate this difference. We present the results of one grid here and discuss the differences between the grids in Appendix~\ref{sec:grids}.

The mean of $(1+\delta)$ is 1, the mean of $e^s$, however, is higher than 1, since the fluctuations are not symmetric. In order to properly compare the resulting spectra we therefore absorb a factor of $\left\langle e^s \right\rangle$ into $\rho_0$, i.e., in the following discussion we compare the power spectra of $\delta$ and $(e^s/\!\left\langle e^s \right\rangle - 1)$.

\subsection{The mildly non-linear regime}
\label{sec:mid-non-linear}
\begin{figure*}
 \centering
 \includegraphics[width=0.49\linewidth]{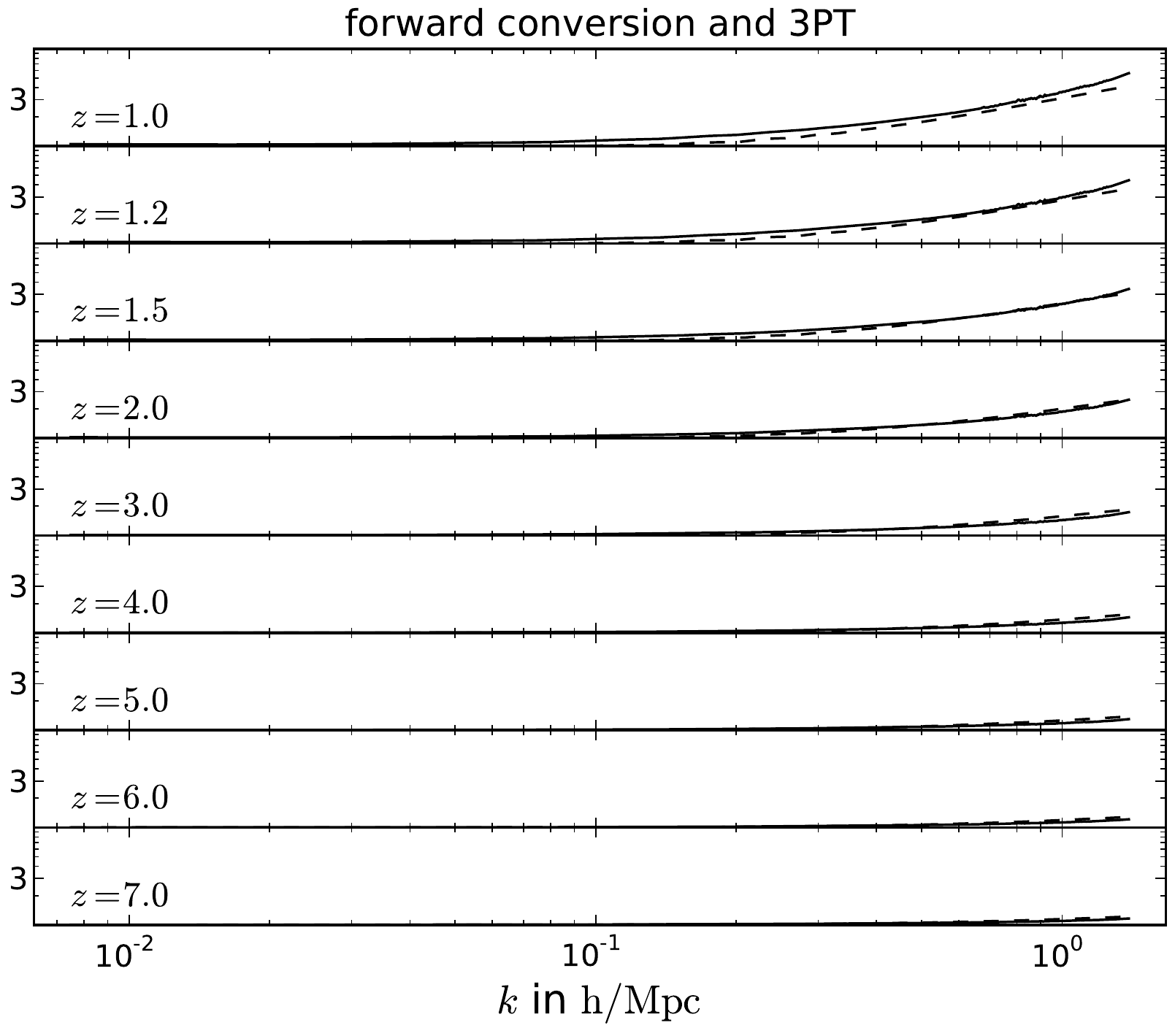}\ \includegraphics[width=0.502\linewidth]{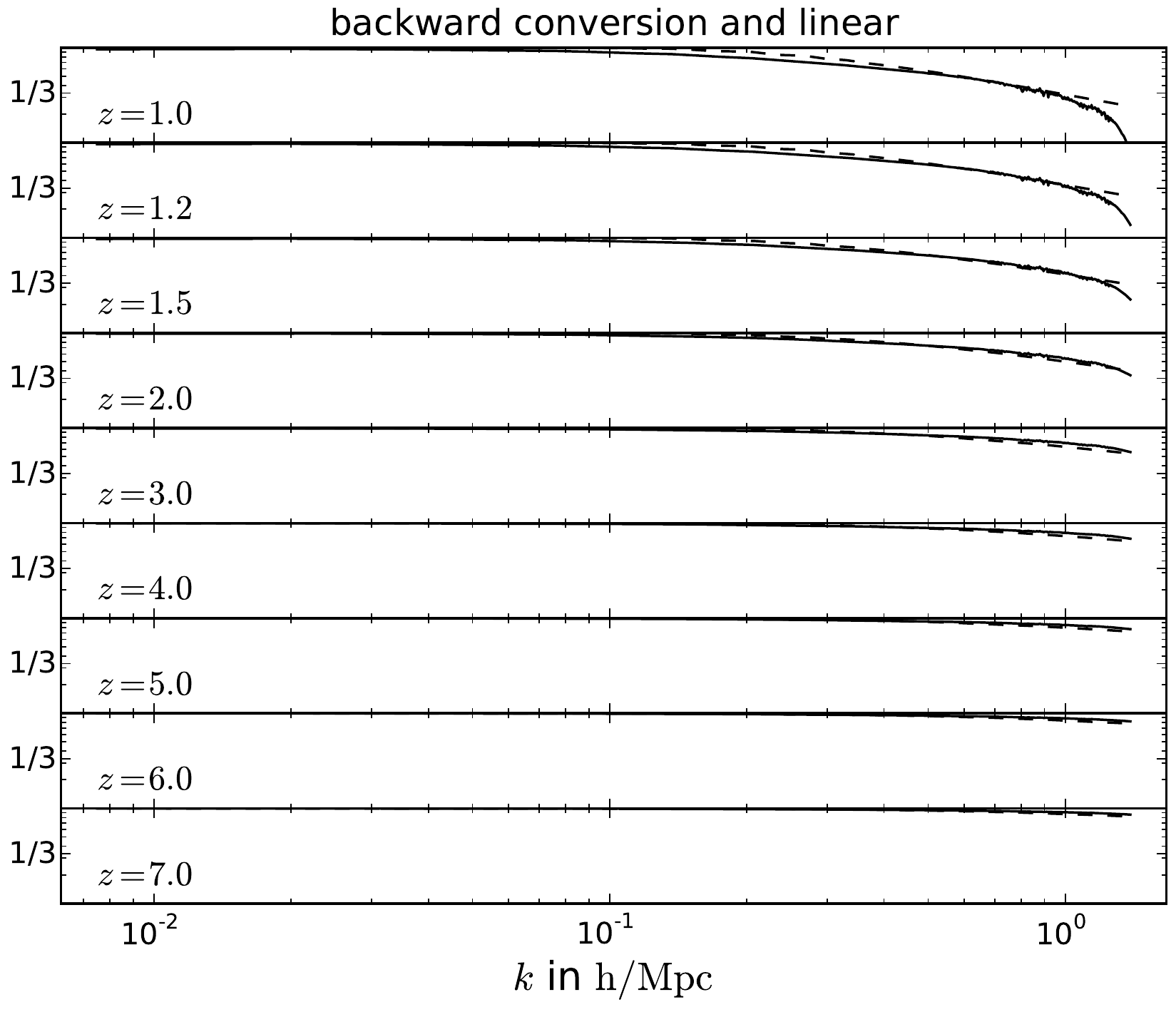}
 \caption{The conversion formalism for redshifts 1 to 7. The left panel shows forward converted linear spectra (solid lines) and the corresponding 3PT spectra (dashed lines), which are both divided by the respective linear spectrum for better comparison of the non-linearities. The right panel shows backward converted 3PT spectra (solid lines) and linear spectra (dashed lines), which are both divided by the respective 3PT spectrum.}
 \label{fig:low_all}
\end{figure*}

\begin{figure*}
 \centering
 \includegraphics[width=0.496\linewidth]{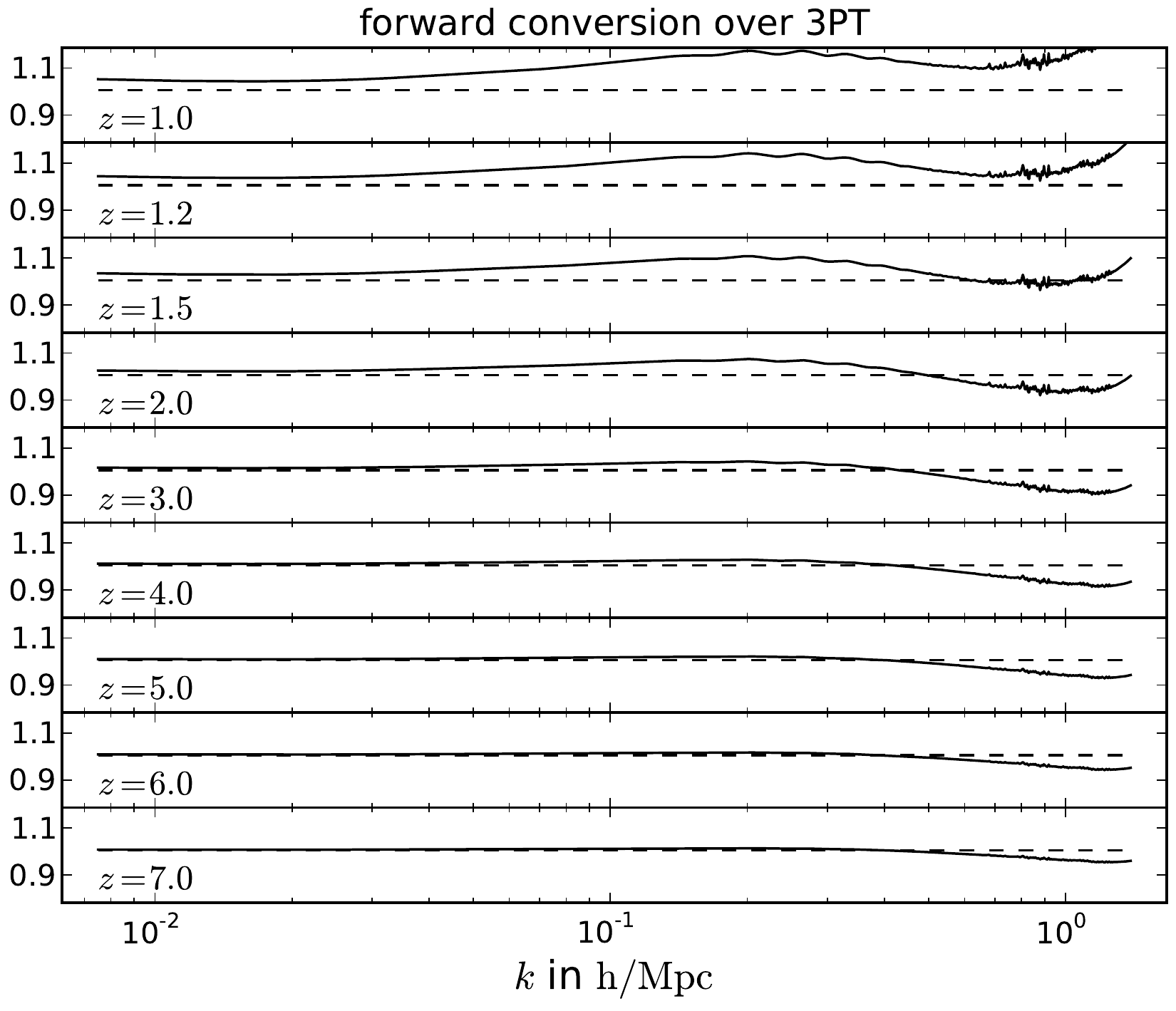}\ \includegraphics[width=0.496\linewidth]{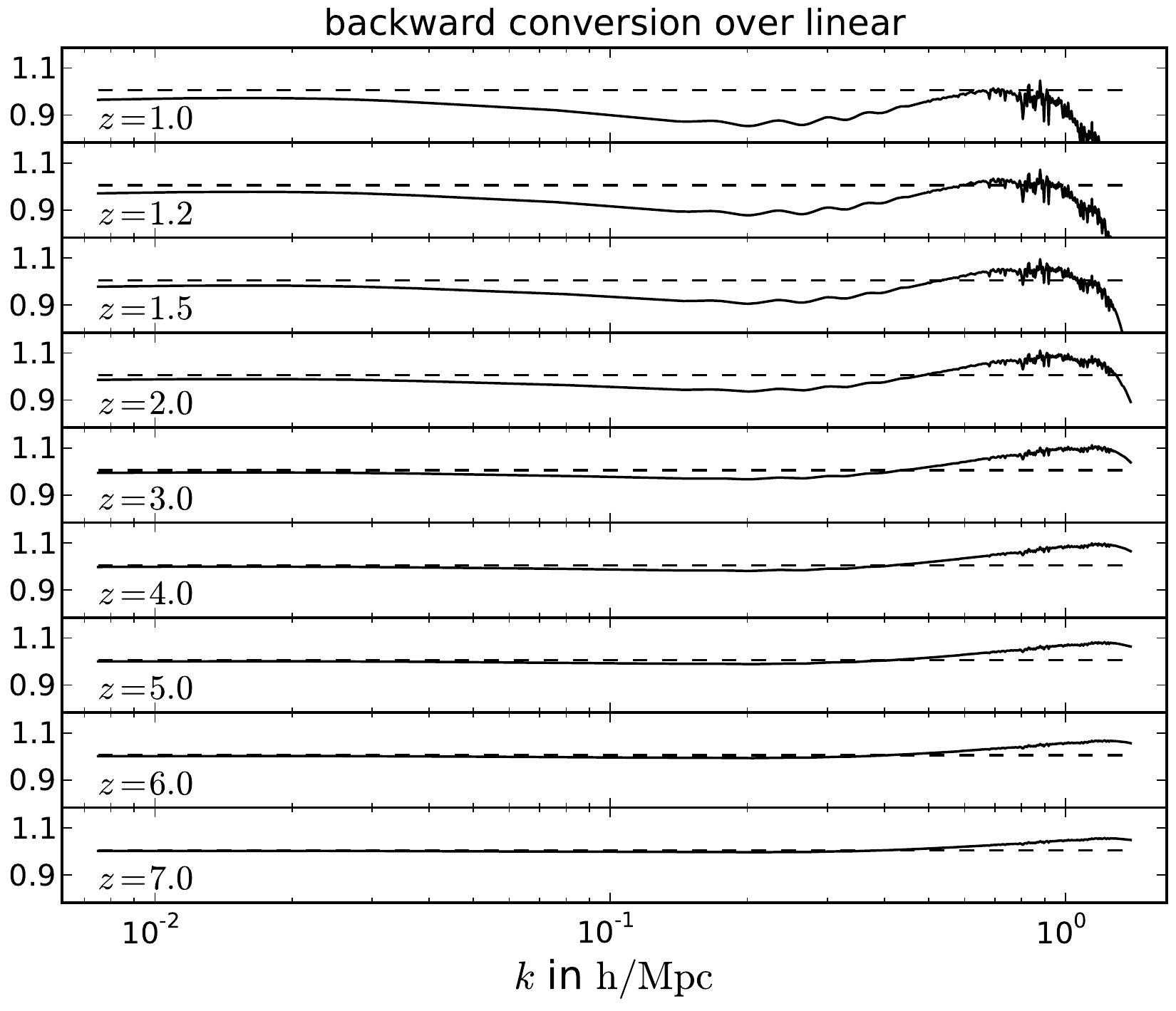}
 \caption{The conversion formalism for redshifts 1 to 7. The left panel shows forward converted linear spectra (solid lines) divided by the corresponding 3PT spectra. The right panel shows backward converted 3PT spectra (solid lines) divided by the respective linear spectrum. The dashed line marks 1.}
 \label{fig:low_over_all}
\end{figure*}

For redshifts higher or equal to 1, the non-linear corrections (calculated using 3PT) are rather mild. In this regime the model of a linearly evolved log density contrast works rather well, as we depict in Fig.~\ref{fig:low_all}.
For a more quantitative comparison we depict the forward converted spectra divided by the 3PT spectra and the backward converted spectra divided by the linear spectra in Fig.~\ref{fig:low_over_all}.
For redshifts 1 to 7 and $k\leq1.0\,h\,\mathrm{Mpc}^{-1}$ the maximal log-distance between the converted spectra and the 3PT spectra stays below $0.17$. The distances are slightly lower for the backward conversion. As one can see in Fig.~\ref{fig:low_over_all}, the distance is strongest around $k\approx 0.2\,h\,\mathrm{Mpc}^{-1}$. For a full list of the log-distances see Appendix~\ref{sec:log-distances}.
In the region where $1.0\,h\,\mathrm{Mpc}^{-1}<k<1.4\,h\,\mathrm{Mpc}^{-1}$ the results are susceptible to the choice of gridding. We discuss this in detail in Appendix~\ref{sec:grids}.

\subsection{The non-linear regime}
\label{sec:non-linear}

\begin{figure*}
 \centering
 \includegraphics[width=0.49\linewidth]{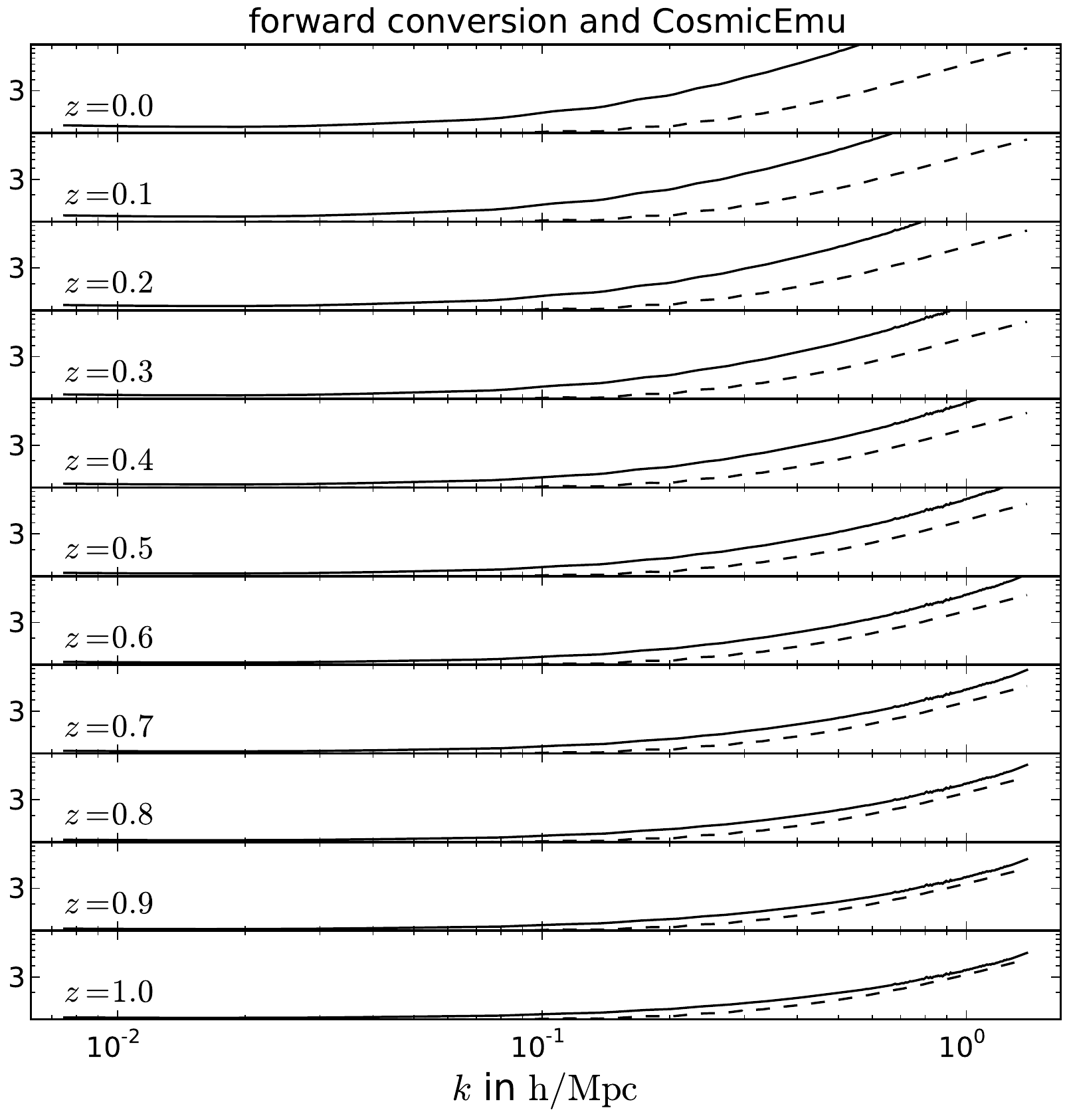}\ \includegraphics[width=0.502\linewidth]{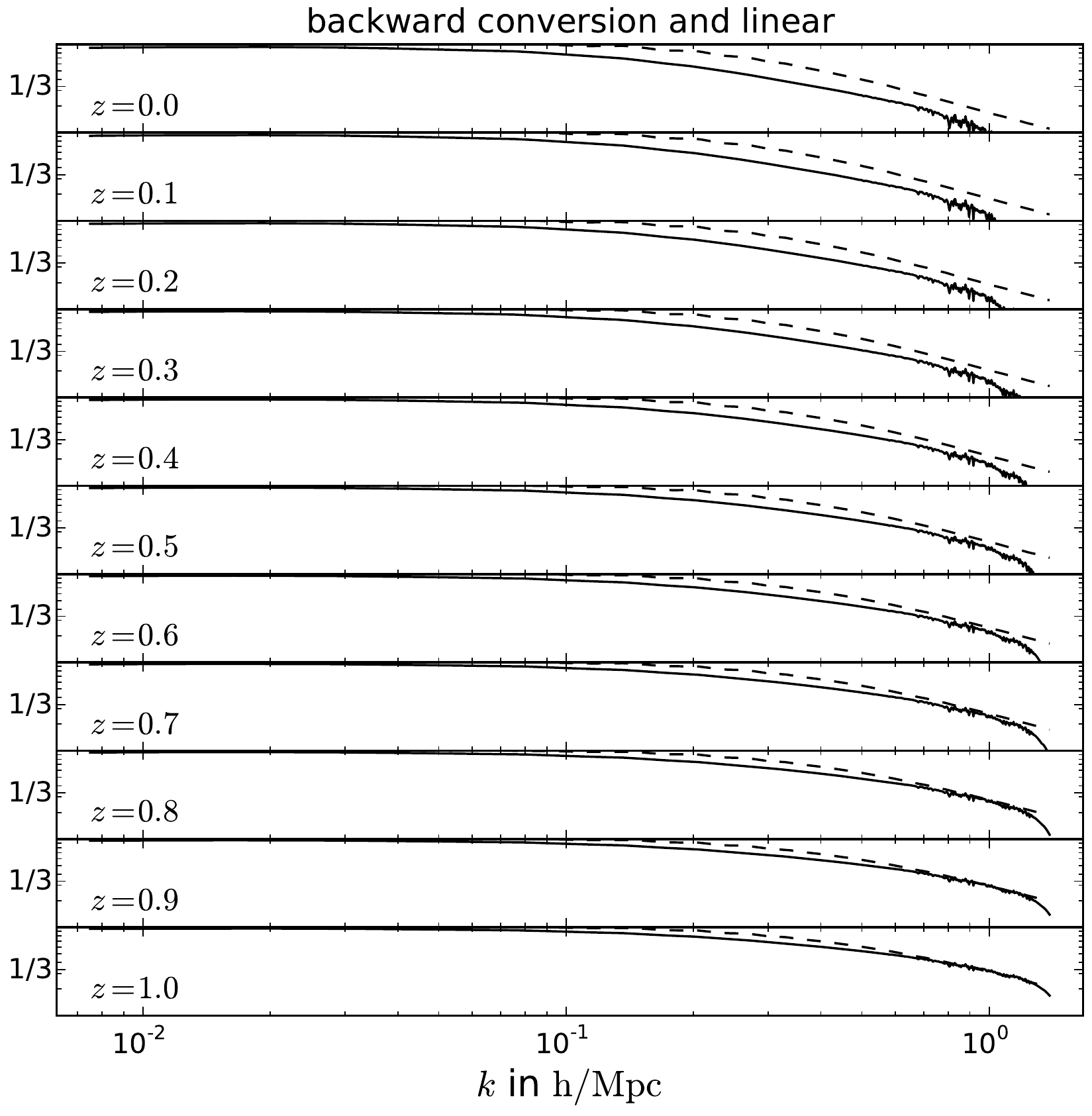}
 \caption{The conversion formalism for redshifts 0 to 1. The left panel shows forward converted linear spectra (solid lines) and the corresponding CosmicEmu spectra (dashed lines), which are both divided by the respective linear spectrum for better comparison of the non-linearities. The right panel shows backward converted CosmicEmu spectra (solid lines) and linear spectra (dashed lines), which are both divided by the respective CosmicEmu spectrum.}
 \label{fig:high_all}
\end{figure*}

\begin{figure*}
 \centering
 \includegraphics[width=0.496\linewidth]{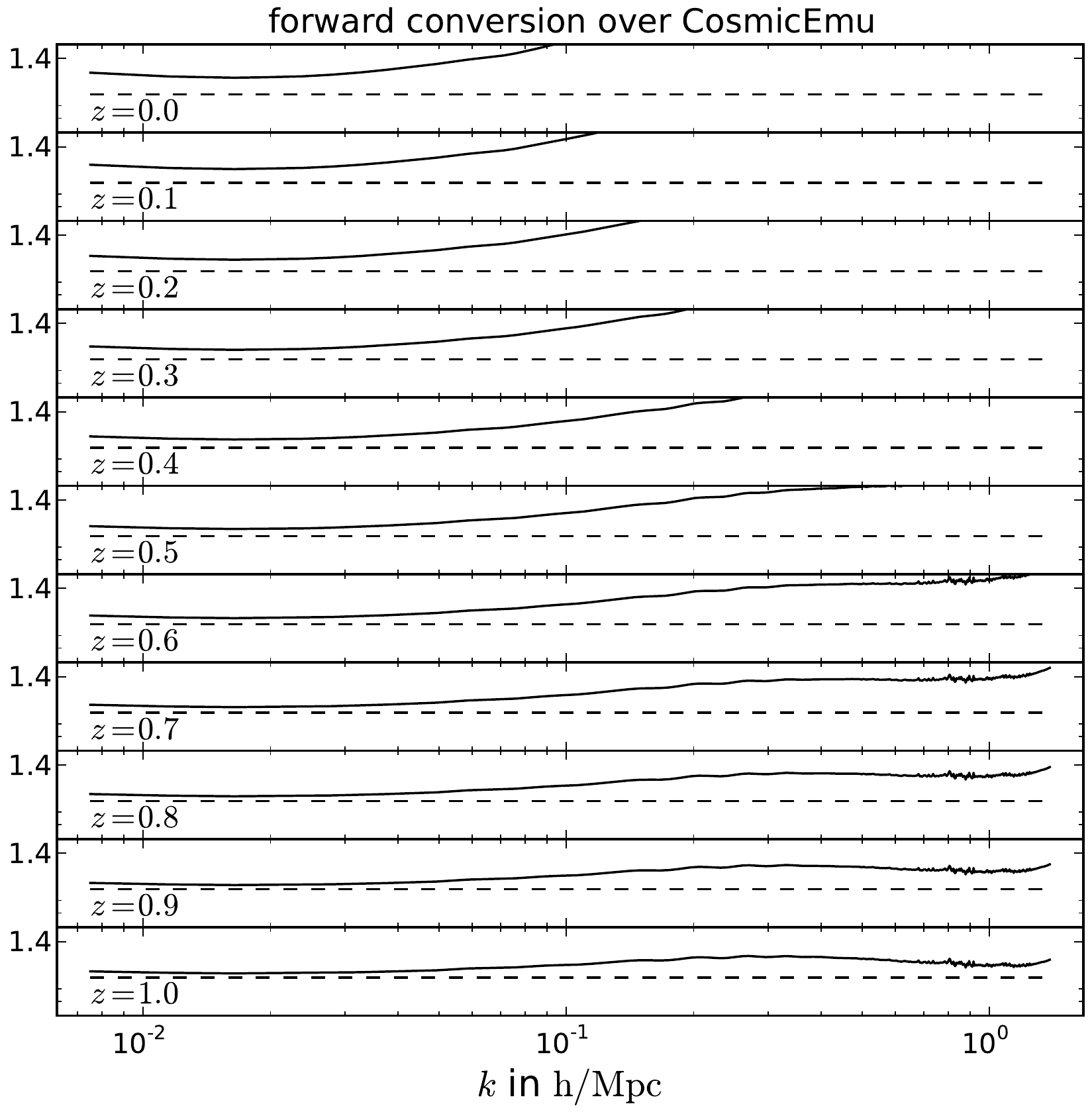}\ \includegraphics[width=0.496\linewidth]{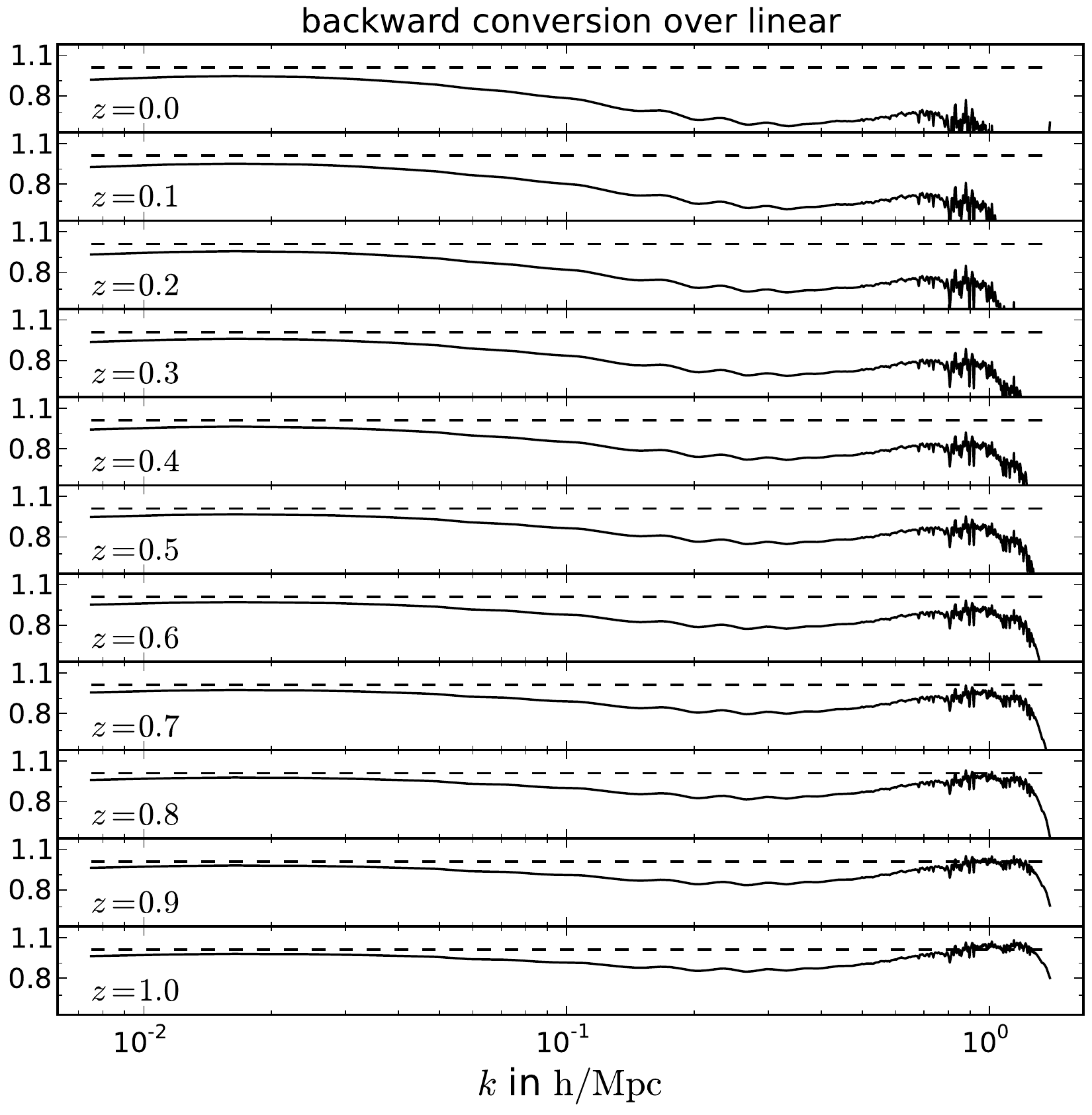}
 \caption{The conversion formalism for redshifts 0 to 1. The left panel shows forward converted linear spectra (solid lines) divided by the corresponding CosmicEmu spectra. The right panel shows backward converted CosmicEmu spectra (solid lines) divided by the respective linear spectrum. The dashed line marks 1.}
 \label{fig:high_over_all}
\end{figure*}

For redshifts smaller than 1, the non-linear corrections add small-scale power which is comparable to the total linear power. On the grid we chose, the total power of the CosmicEmu spectrum is twice as high as the power of the linear spectrum at redshift $0.8$. In this regime the agreement between the converted spectra and the CosmicEmu (or linear spectra, respectively) is much weaker. We depict the non-linearities generated by the forward conversion and the reduction of non-linearities in the backward conversion in Fig.~\ref{fig:high_all}. A more error focused comparison with the forward converted spectra divided by the CosmicEmu spectra and with the backward converted spectra divided by the linear spectra can be found in Fig.~\ref{fig:high_over_all}.
The maximal log-distance for $k\leq1.0\,h\,\mathrm{Mpc}^{-1}$ between the backward converted spectra and the CosmicEmu spectra ranges between $0.17$ for $z=1$ and $0.8$ for $z=0$. For the forward converted spectra it ranges from $0.2$ at $z=1$ to $1.5$ $z=0$. Clearly, a log-distance of $1.5$(corresponding to a factor of $4.5$) is not within an acceptable margin of error. Emulating the non-linear corrections from CosmicEmu by forward conversion breaks down in this regime. The backward conversion performs better than the forward conversion, but a log-distance of $0.8$(corresponding to a factor of $2.2$) is far from ideal. However, the backward conversion still reduces the magnitude of non-linear corrections significantly. See Table~\ref{table:log-distances} in Appendix~\ref{sec:log-distances} for a detailed comparison of the log-distances.

\subsection{Smearing of baryon acoustic oscillations}
\begin{figure*}
 \centering
 \includegraphics[width=0.49\linewidth]{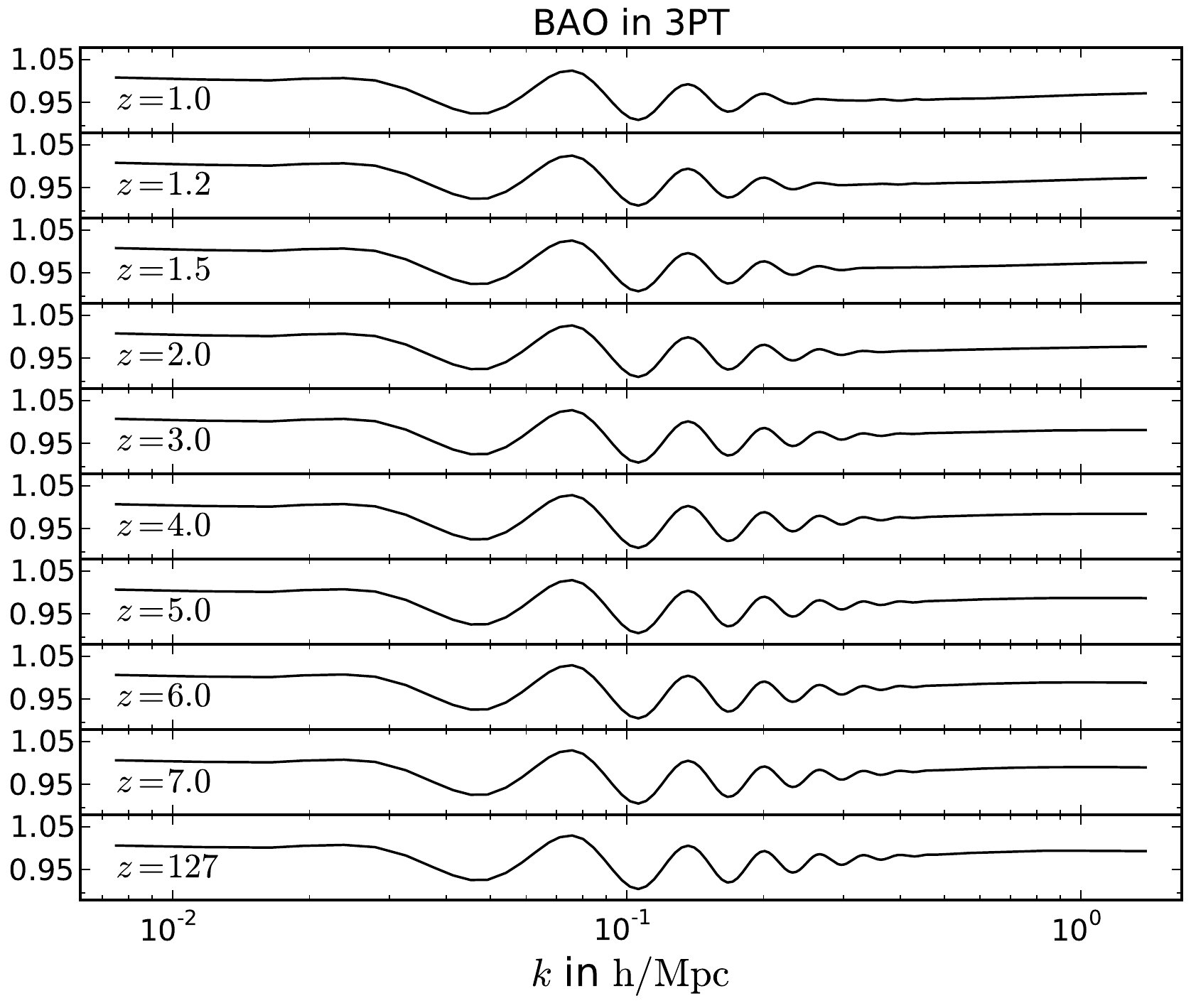}\ \includegraphics[width=0.49\linewidth]{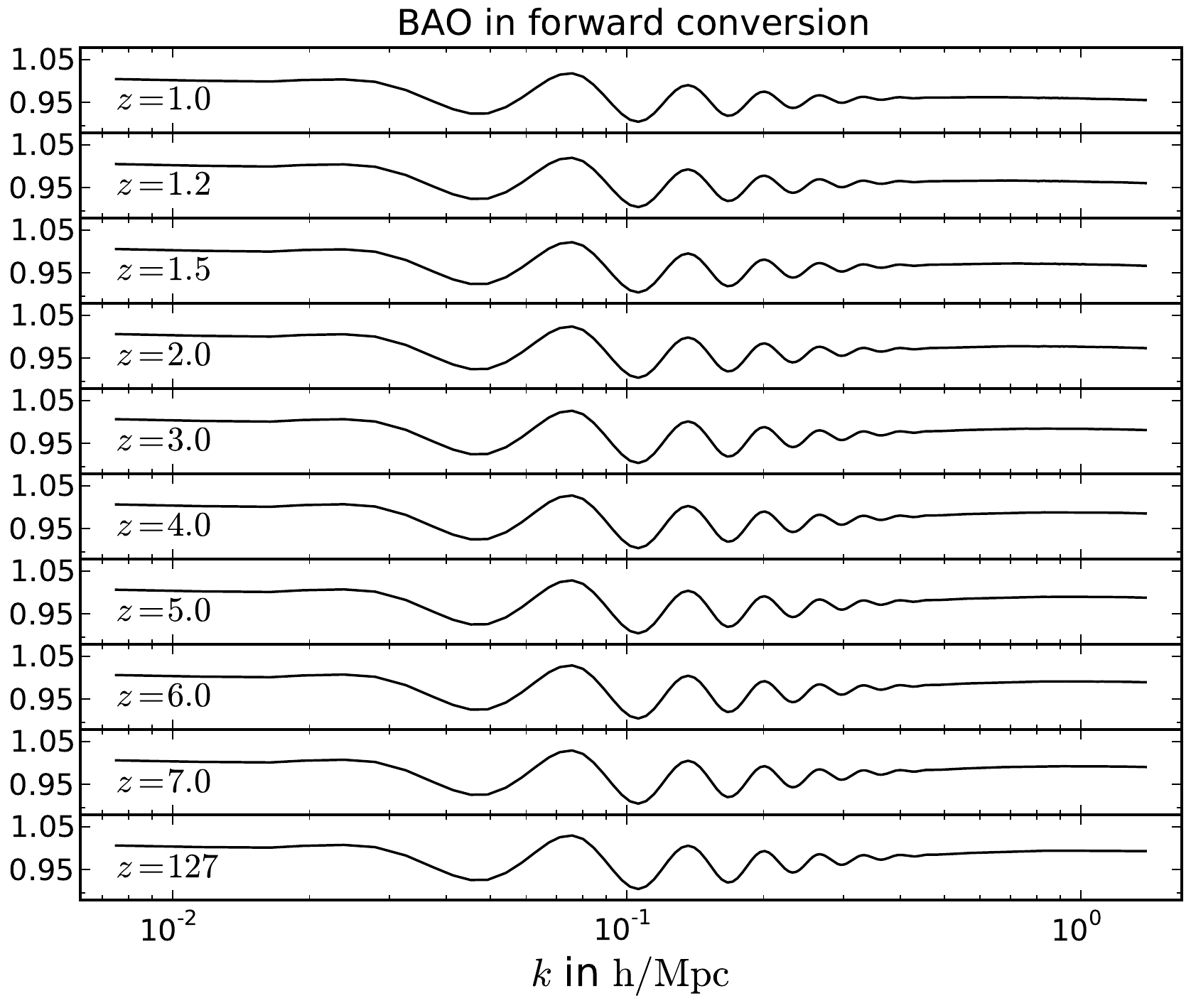}
 \caption{BAO smearing. The left panel shows the 3PT spectra with BAO divided by the 3PT spectra without BAO at several redshifts. The right panel shows the forward converted linear spectra with BAO divided by the forward converted linear spectra without BAO. We note the smearing of small-scale wiggles towards lower redshifts in the left panel.}
 \label{fig:low_BAO}
\end{figure*}

Another important non-linear effect is the smearing of baryon acoustic oscillations (BAO). Towards lower redshifts the smallest scale wiggles in the matter power spectrum are erased while the large scale wiggles become increasingly damped as redshift decreases. This can be nicely seen in Fig.~\ref{fig:low_BAO}, where we depict a comparison between 3PT spectra calculated on the basis of the CAMB power spectrum and 3PT spectra calculated on the basis of the ``no wiggle'' power spectrum by \cite{Eisenstein-1998}\footnote{The ``no wiggle'' power spectrum was calculated by putting the cosmological parameters from \cite{Planck-2013} into the code from \url{http://www.mpa-garching.mpg.de/~komatsu/CRL/powerspectrum/nowiggle/}.} in the left panel. The same comparison but using the forward converted linear spectra instead of the 3PT spectra can be seen in the right panel. It is evident that BAO smearing appears only very slightly in the forward converted power spectra, it is much weaker than the smearing calculated using 3PT. BAO smearing appears to be about 3 to 4 times weaker (see Appendix~\ref{sec:smear_strength}) in the forward converted spectra compared to the 3PT spectra, which exhibit a smearing in agreement with N-body simulations.

\section{Discussion and conclusions}
\label{sec:conclusions}

Starting from the assumption of Gaussian statistics for $s$ we presented a formalism to calculate the power spectrum of $s=\log(\rho)$ from the power spectrum of $\rho$ and vice versa. This formalism can for example be used to compare theoretical considerations about logarithmic densities with theoretical or observed power spectra, which are calculated from the linear density. A numerical implementation of these formulas can be found in the \textsc{NIFTy} python package\footnote{\url{http://www.mpa-garching.mpg.de/ift/nifty/}} (\cite{Selig-2013}). The formalism is also useful to generate log-normal fields with a given power spectrum, as one can simply convert the spectrum, use the resulting spectrum to generate Gaussian random fields, and then exponentiate them (see Appendix~\ref{sec:generate_lognormal}).

We used this formalism to investigate whether the non-linear corrections to the matter spectrum are reduced for the logarithmic density and whether the non-linearities can be emulated by applying the growth factor to the log-density instead of the density. To that end we compared emulated linear (CAMB) and non-linear (third-order perturbation theory and CosmicEmu) spectra with the in- and outputs of our formulas. For the mildly non-linear regime (redshift 1-7) we find that the log-transformed non-linear spectra agree with the linear spectra rather well (less than 20\% difference between them up to $k=1.0\,h\,\mathrm{Mpc}^{-1}$). This enables one to easily generate a log-normal field, which follows the appropriate non-linear power spectrum, even for position dependent redshifts. We describe this procedure in detail in Appendix~\ref{sec:generate_matter}. For lower redshifts the agreement decreases but the non-linearities are still reduced significantly. The smearing of baryon acoustic oscillations can not be emulated by applying the growth factor to the log-density. There is some smearing due to mode coupling, but it is about 3 to 4 times weaker than in the more exact third-order perturbation theory.

We performed the calculation on four different grids leading to different but within the achieved precision comparable results. The differences between the results on different grids indicate that the agreement between our model and the non-linear spectra from literature decreases if the grid allows scales corresponding to $k\geq 1.4\,h\,\mathrm{Mpc}^{-1}$, since mode coupling to the small scales overestimates the non-linear corrections. This problem arises since the total power of the matter spectrum diverges as its spectral index is above $-3$, which makes some sort of cut-off in power necessary in a non-linear transformation such as the exponential and logarithmic functions. Here, this cut-off is imposed by the resolution of the grid. Ultimately, one would have to find a physically justifiable way to regularize the integral in Eq.~\ref{eq:angles_integrated}.

\begin{acknowledgements}
 The authors would like to thank Niels Oppermann, Philipp Wullstein, Sebastian Dorn, and Marco Selig for fruitful discussions and support. We would also like to thank the anonymous referee for the constructive review and useful comments. All numerical calculations were done using the \textsc{NIFTy} python package by \cite{Selig-2013}.
\end{acknowledgements}

\bibliographystyle{aa}
\bibliography{powspec}

\begin{appendix}

\section{Generating log-normal fields from a given power spectrum}
\label{sec:generate_lognormal}

We can use the backward conversion (Eqs.~\eqref{eq:backward_no_mono}~\&~\eqref{eq:backward_mono}) to generate log-normal random fields that follow a given linear power spectrum.
This method has already been presented by \cite{Percival-2004} (see Sect.~3.2 therein). For illustrative purposes we demonstrate the procedure for an isotropic case, but the same procedure can be applied for an anisotropic power spectrum.

Suppose the power spectrum $P_\rho(k)$ is known within some range $k_\mathrm{min}<k<k_\mathrm{max}$ and the mean $\left\langle\rho\right\rangle$ is also known.
First, we set up a discretized space in which $k_\mathrm{min}$ corresponds to the minimal non-zero mode and $k_\mathrm{max}$ is the highest supported mode.
We construct the monopole according to Eq.~\eqref{eq:fill_mono} and apply Eqs.~\eqref{eq:backward_no_mono}~and~\eqref{eq:backward_mono} to the power spectrum. Setting $P_s(k\!=\!0)=0$ in Eq.~\eqref{eq:backward_mono} we end up with a power spectrum $P_s(k)$ and a mean $m$. We can now construct log-normal fields by generating Gaussian random fields from $P_s(k)$ and exponentiating the sum of $s$ and $m$. The resulting field \mbox{$\rho=e^{s+m}$} follows (in the statistical average) the power spectrum $P_\rho(k)$.

\begin{figure}
 \centering
 \includegraphics[width=\linewidth]{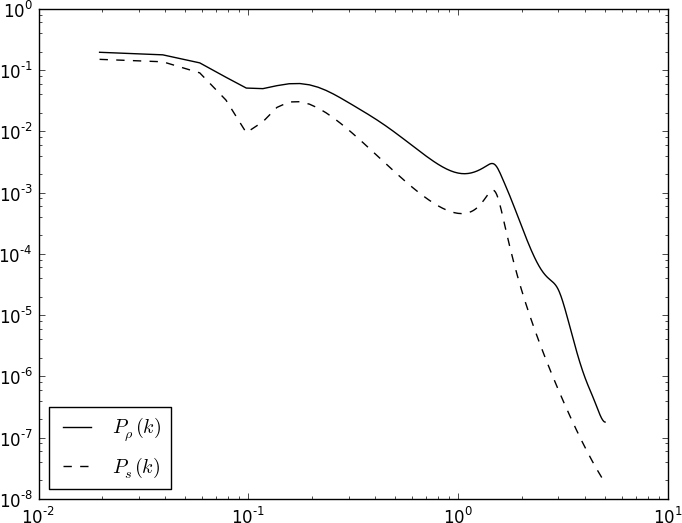}
 \caption{The backward conversion applied to a one-dimensional power spectrum $P_\rho(k)$ (solid line) yields the power spectrum $P_s(k)$ (dashed line).}
 \label{fig:mock_conv}
\end{figure}

\begin{figure}
 \centering
 \includegraphics[width=0.952\linewidth]{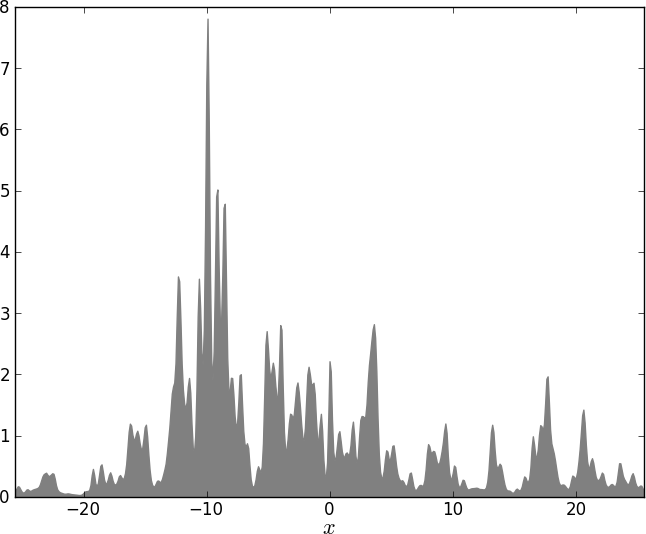}
 \caption{A one-dimensional log-normal random field following the power spectrum $P_\rho(k)$ from Fig.~\ref{fig:mock_conv}.}
 \label{fig:mock_rho}
\end{figure}

We illustrate this in Fig.~\ref{fig:mock_conv}. Starting with the power spectrum of $\rho$ and $\left\langle\rho\right\rangle=1$, we calculate the power spectrum and mean of $s$. The mean is $-0.76$. With these quantities we can draw log-normal fields that follow the original power spectrum. One random field of that kind is plotted in Fig.~\ref{fig:mock_rho}.

\section{The discretized conversion formula}
\label{sec:discrete}
The discetized conversion formulas are implemented in the \textsc{NIFTy} python package by \cite{Selig-2013} (version 0.8.4 or higher).

In discretizing the forward and backward conversion formulas from Sect.~\ref{sec:forward}~and~\ref{sec:backward} one needs to pay careful attention to volume factors arising from the discretization of the integrals. We define the discrete Fourier transform as
\begin{equation}
 s(\slashed{\vec{k}}) = \sum\limits_{\vec{x}} V_x \, e^{i 2\pi\, \slashed{\vec{k}}\cdot\vec{x}} s(\vec{x})
\end{equation}
We note that the definition of the Fourier vector has changed with respect to the main text, $\slashed{\vec{k}} = \vec{k}/(2\pi)$, as this is more common in a numerical setting.
The vector $\vec{x}$ can only take discrete values,
\begin{equation}
\begin{split}
 \vec{x}_j & = n_j \Delta x_j \quad \mathrm{with}\\
 n_j & \in \{ -N_j/2,\ -N_j/2+1,\ ...\ ,\ N_j/2-1\},
\end{split}
\end{equation}
where the dimension $j$ is discretized into $N_j$ pixels with edge length $\Delta x_j$.
Therefore, our discretized space consists of $N = \prod_j N_j$ pixels that have the volume
\begin{equation}
 V_x = \prod_j \Delta x_j.
\end{equation}
This defines a conjugate set of discrete values for $\slashed{\vec{k}}$,
\begin{equation}
\begin{split}
 \slashed{\vec{k}}_j & = n_j \Delta \slashed{k}_j \quad \mathrm{with} \\
 n_j & \in \{ -N_j/2,\ -N_j/2+1,\ ...\ ,\ N_j/2-1\},
\end{split}
\end{equation}
where $\Delta \slashed{k}_j = 1/(N_j\,\Delta x_j)$.
Therefore, the conjugate space consists of $N$ pixels that have the volume $V_k = 1/(N\, V_x)$.
The power spectrum is defined analogously to Eq.~\eqref{eq:first_powspec} by
\begin{equation}
 \sum\limits_{\vec{x}\,\vec{y}} e^{i 2\pi\, \slashed{\vec{k}}\cdot\vec{x} - i 2\pi\, \slashed{\vec{q}}\cdot\vec{y}}\ \left\langle s(\vec{x}) s(\vec{y}) \right\rangle_{\mathcal{P}(s)} = \frac{1}{V_k} \delta_{\slashed{\vec{k}}\slashed{\vec{q}}}\, P_s(\slashed{\vec{k}}),
\end{equation}
where
\begin{equation}
 \delta_{\slashed{\vec{k}}\slashed{\vec{q}}}= \begin{cases}
                          1\quad \mathrm{if}\quad \slashed{\vec{k}}=\slashed{\vec{q}}\\
			  0\quad \mathrm{otherwise}
                         \end{cases}.
\end{equation}
Defining $\rho(\vec{x}) := e^{m+s(\vec{x})}$ one can derive\footnote{Assuming Gaussianity and statistical homogeneity of $s$.} the forward conversion formula as
\begin{equation}
 P_\rho(\slashed{\vec{k}}) = \sum\limits_{\vec{x}}V_x\, e^{i 2\pi\, \slashed{\vec{k}}\cdot\vec{x}}\, e^{2m} \exp\!\left( \sum\limits_{\slashed{\vec{q}}}V_k\! \left(e^{-i 2\pi\, \vec{x}\cdot\slashed{\vec{q}}}+1\right) P_s(\slashed{\vec{q}}) \right).
\end{equation}
The backward conversion formula is
\begin{equation}
\begin{split}
   P_s(\slashed{\vec{k}}\neq\vec{0}) & = \sum\limits_{\vec{x}}V_x \ e^{i2\pi\, \slashed{\vec{k}}\cdot\vec{x}}\ \log\!\left( \sum\limits_{\slashed{\vec{q}}}V_k \, e^{-i2\pi\, \slashed{\vec{q}}\cdot\vec{x}}\ P_\rho(\slashed{\vec{q}})\right)
\end{split}
\end{equation}
and for the monopole
\begin{equation}
\begin{split}
 P_s(\slashed{\vec{k}}=\vec{0}) & = \sum\limits_{\vec{x}}V_x \ \log\!\left( \sum\limits_{\slashed{\vec{q}}}V_k \, e^{-i2\pi\, \slashed{\vec{q}}\cdot\vec{x}}\ P_\rho(\slashed{\vec{q}})\right)\\
  &\quad - \frac{1}{2}\frac{1}{V_k} \log\!\left( \sum\limits_{\slashed{\vec{q}}}V_k \, P_\rho(\slashed{\vec{q}}) \right)
 - \frac{1}{V_k}  m.
\end{split}
\end{equation}

\section{The conversion for spherical harmonics}
\label{sec:spherical}
The spherical harmonics conversion formulas are implemented in the \textsc{NIFTy} python package by \cite{Selig-2013} (version 0.8.4 or higher).

We define the spherical harmonics as
\begin{equation}
Y_l^m( \theta , \varphi ) = \sqrt{\frac{(2l+1)}{4\pi}\frac{(l-m)!}{(l+m)!}}  \, P_l^m ( \cos{\theta} ) \, e^{i m \varphi },
\end{equation}
where we will write $\Omega$ as a short-hand notation for $\theta , \varphi$ with $\Omega=0$ corresponding to $\theta =0$, $\varphi = 0$.
In the spherical harmonics basis we define the power spectrum as
\begin{equation}
 P_s(l)\ \delta_{ll'} \delta_{mm'} = S^{l'm'}_{lm} = \left\langle s_{lm} s^*_{l'm'} \right\rangle_{\mathcal{P}(s)},
\end{equation}
where we assumed statistical homogeneity and isotropy. In this section, $\delta_{ij}$ denotes the Kronecker delta and $^*$ denotes complex conjugation.
The covariance matrix in pixel space is related to the power spectrum by
\begin{equation}
\begin{split}
 S(\Omega,\Omega')  = \sum\limits_{l=0}^\infty \sum\limits_{m=-l}^l \sum\limits_{l'=0}^\infty \sum\limits_{m=-l'}^{l'}\ & Y^m_l(\Omega)\, Y^{m'}_{l'}\!(\Omega')\\
 & \times P_s(l)\ \delta_{ll'} \delta_{mm'},\\
\end{split}
\end{equation}
which in the case of $\Omega'=0$ is 
\begin{equation}
 S(\Omega,0)  = \sum\limits_{l=0}^\infty \sum\limits_{m=-l}^l\ Y^m_l(\Omega)\ \delta_{m0}\, \sqrt{\frac{2l+1}{4\pi}}\, P_s(l).
\end{equation}
Because of statistical homogeneity and isotropy, this stays the same for all angles that are separated by $\Omega$.

Following a similar calculation as in Sects.~\ref{sec:forward} and \ref{sec:backward} we derive the forward conversion as
\begin{equation}
\begin{split}
 P_\rho(l) =\ & \sqrt{\frac{4\pi}{2l+1}} \int\!\!\mathrm{d}\Omega\ Y^{0}_l(\Omega)^*\, e^{2m} \\ 
 & \times \exp\!\left( \sum\limits_{l'=0}^\infty \left( Y^0_{l'}(\Omega) +\sqrt{\frac{2l'+1}{4\pi}} \right) \sqrt{\frac{2l'+1}{4\pi}} P_s(l')\right)
\end{split}
\end{equation}
and the backward conversion as
\begin{equation}
\begin{split}
& P_s(l\!\neq\!0) = \\
& \sqrt{\frac{4\pi}{2l+1}} \int\!\!\mathrm{d}\Omega\ Y^{0}_{l}\!(\Omega)^*\,\log\!\left( \sum\limits_{l'=0}^\infty\ Y^0_{l'}(\Omega)\, \sqrt{\frac{2l'+1}{4\pi}}\, P_\rho(l') \right) \\
\end{split}
\end{equation}
and
\begin{equation}
\begin{split}
& P_s(l\!=\!0) + 4\pi\,m = \\
 & \int\!\!\mathrm{d}\Omega\ \log\!\left( \sum\limits_{l'=0}^\infty\ Y^0_l(\Omega)\, \sqrt{\frac{2l+1}{4\pi}}\, P_\rho(l) \right) \\
 & \left. -\frac{1}{2}\sqrt{4\pi}\, \log\!\left( \sum\limits_{l'=0}^\infty\ \frac{2l'+1}{4\pi}\, P_\rho(l') \right) \right].
\end{split}
\end{equation}

\section{Supplements to Sect.~\ref{sec:matter_spectrum}}
\label{sec:matter_appendix}
\subsection{Log-distances}
\label{sec:log-distances}

We calculate the maximal log-distance between two spectra as
\begin{equation}
 \mathrm{max}_k \left| \log(P_1(k)) - \log(P_2(k)) \right|.
\end{equation}
This distance gives a quantitative measure about the goodness of approximating the non-linear spectra by forward converted spectra and about the reduction of non-linearities in the backward converted spectra. In Sects.~\ref{sec:mid-non-linear}~\&~\ref{sec:non-linear} we give an overview over the log-distances for $k\leq1.0\,h\,\mathrm{Mpc}$. Here, we list all of their values in Table~\ref{table:log-distances}, which also shows the log-distances between the linear and non-linear spectra. As one can see the converted spectra always reduce the log-distance.

\begin{table}
\caption{Highest log-distance (for $k\leq1.0\,h\,\mathrm{Mpc}^{-1}$) between the forward converted spectra and the emulated spectra (3PT and CosmicEmu, respectively) and the backward converted emulated spectra and the linear spectra. We also list the log-distance between the linear spectra and the emulated spectra for comparison.}              
\label{table:log-distances}      
\centering                                      
\begin{tabular}{c c c c}          
\hline\hline
   redshift & forward \& & backward \& & emulated \& \\    
    & emulated & linear & linear \\
\hline                                   
 7 & 0.04 & 0.04 & 0.14\\
 6 & 0.05 & 0.05 & 0.17 \\
 5 & 0.07 & 0.07 & 0.23\\
 4 & 0.08 & 0.08 & 0.31\\
 3 & 0.09 & 0.10 & 0.45\\
 2 & 0.09 & 0.11 & 0.69\\
 1.5 & 0.10 & 0.10 & 0.88\\
 1.2 & 0.14 & 0.13 & 1.0\\
 1.0 & 0.17 (3PT) & 0.15 (3PT) & 1.1 (3PT)\\
  & 0.2 (Emu) & 0.17 (Emu) & 1.2 (Emu)\\
 0.9 & 0.2 & 0.19 & 1.2\\
 0.8 & 0.3 & 0.2 & 1.3\\
 0.7 & 0.4 & 0.2 & 1.3\\
 0.6 & 0.4 & 0.3 & 1.4\\
 0.5 & 0.6 & 0.3 & 1.5\\
 0.4 & 0.7 & 0.4 & 1.5\\
 0.3 & 0.8 & 0.4 & 1.6\\
 0.2 & 1.0 & 0.5 & 1.7\\
 0.1 & 1.3 & 0.7 & 1.7\\
 0.0 & 1.5 & 0.8 & 1.8\\
\hline
\end{tabular}
\end{table}

\subsection{The effect of different grids}
\label{sec:grids}

The calculation in Sect.~\ref{sec:matter_spectrum} is performed on four different grids which all cover the same physical volume, but with different resolution. The Figures in Sect.~\ref{sec:matter_spectrum} show the results of the second grid (grid B).
The lowest non-zero mode we want to cover is $k=0.0075\,h\,\mathrm{Mpc}^{-1}$ for all grids. The highest mode is different for the three grids, the coarsest one having $k_\mathrm{max}=1.0\,h\,\mathrm{Mpc}^{-1}$.
The minimum spectral length covered by a three-dimensional Cartesian box with the origin in the middle is
\begin{equation}
k_\mathrm{min} = 2\pi\left(\mathrm{max}\left\lbrace L_x,L_y,L_z\right\rbrace\right)^{-1},
\end{equation}
where $L_x$, $L_y$, and $L_z$ are the total edge lengths of the box. The maximum spectral length is
\begin{equation}
k_\mathrm{max} = \pi\sqrt{\Delta x^{-2}+\Delta y^{-2}+\Delta z^{-2}},
\end{equation}
where $\Delta x$, $\Delta y$, and $\Delta z$ are the edge lengths of one pixel.
This leaves us with an infinite number of possible pixelizations. We restrict ourselves to cubic pixels and an equal number of pixels in each dimension leaving us with only two parameters: the number of pixels per dimension $N_\mathrm{pix}$ and the pixel edge length $\Delta x$.

The first grid (grid A) consists of $152^3$, the second grid (grid B) of $214^3$, the third grid (grid C) of $522^3$, and the fourth grid (grid D) of $766^3$ cubic pixels. The pixel edge lengths are chosen in a way that the lowest non-zeros value of $k$ is $0.0075\,h\,\mathrm{Mpc}^{-1}$. We summarize the properties of these grids in Table~\ref{table:grids}. The upper limit of the spectral range of grid D is higher than $1.4\,h\,\mathrm{Mpc}^{-1}$. To investigate the effect of the pixel size on the conversion we cut the power spectrum at $k=1.4\,h\,\mathrm{Mpc}^{-1}$ setting all higher modes to zero before the conversion. Therefore, any difference in the results of grid B and D originates from the different choice of pixel sizes, and because grid D covers a sphere in Fourier space whereas grid B covers a cube, since no individual component of the wavevector in grid B can have a value above $0.8 \,h\,\mathrm{Mpc}^{-1}$ and wavevectors with a length above that can only be reached in the corners of the cube.

\begin{table}
\caption{Grids used for conversion of matter density spectra. Units of $\Delta x$ are $\mathrm{Mpc}/h$, units of $k$ are $h\,\mathrm{Mpc}^{-1}$. For grid D the power spectra are cut before the conversion at $k=1.4\,h\,\mathrm{Mpc}^{-1}$ (setting all higher modes to zero).}              
\label{table:grids}      
\centering                                      
\begin{tabular}{c  c  c  c  c }          
\hline\hline
  & grid A & grid B & grid C & grid D\\    
\hline                                   
 $N_\mathrm{pix}$ & $152^3$ & $214^3$ & $522^3$ & $766^3$ \\      
 $\Delta x$ & $5.51$ & $3.91$ & $1.60$ & $1.09$ \\
   $k_\mathrm{min}$ & $0.0075$& $0.0075$& $0.0075$& $0.0075$ \\
   $k_\mathrm{max}$ & $1.0$ & $1.4$ & $3.4$ & $5.0$, cut at $1.4$ \\
\hline                                             
\end{tabular}
\end{table}

\begin{figure}
 \centering
 \includegraphics[width=\linewidth]{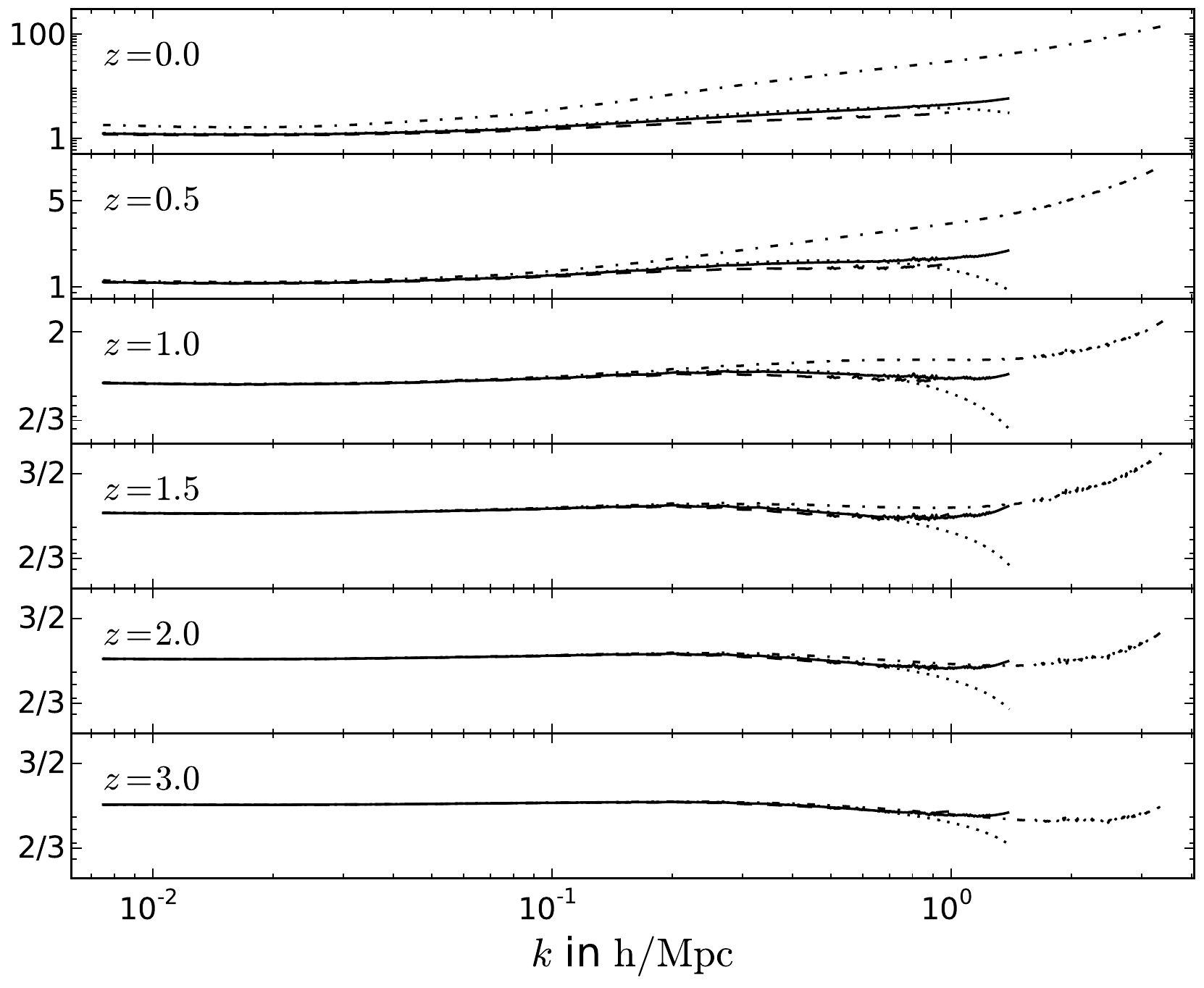}
 \caption{The forward converted spectra using different grids at a selection of redshifts. The panels show the ratio between the forward converted spectrum and the emulated spectrum (CosmicEmu for redshifts 0, 0.5, and 1 and 3PT else). The solid line corresponds to grid B (the grid used in Sect.~\ref{sec:matter_spectrum}), the dashed line to grid A, the dot-dashed line to grid C, and the dotted line to grid D.}
 \label{fig:cut_comp_forward}
\end{figure}

\begin{figure}
 \centering
 \includegraphics[width=\linewidth]{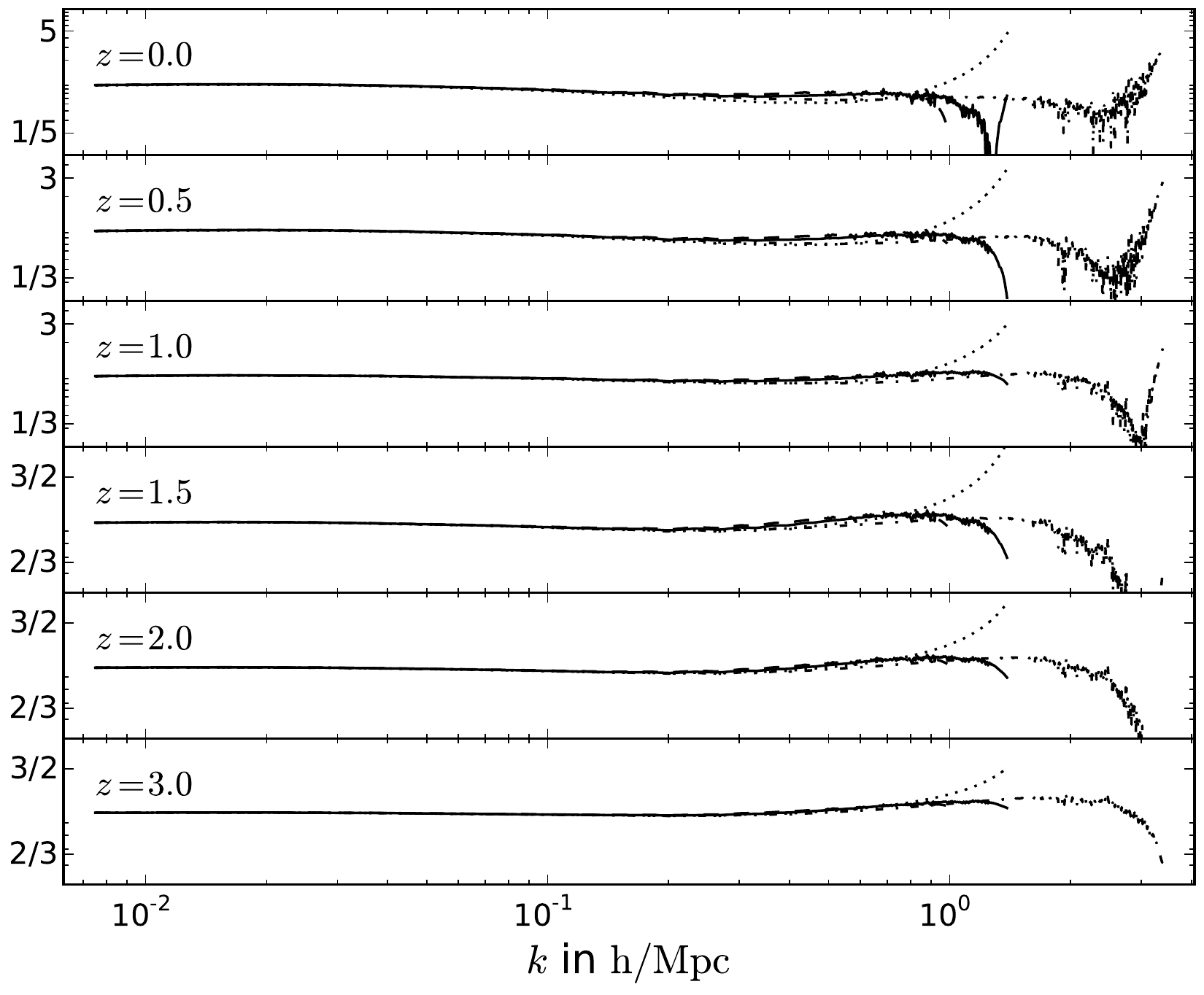}
 \caption{The backward converted spectra using different grids at a selection of redshifts. The panels show the ratio between the backward converted emulated spectrum(emulators as in Fig.~\ref{fig:cut_comp_forward}) and the linear spectrum. The solid line corresponds to grid B (the grid used in Sect.~\ref{sec:matter_spectrum}), the dashed line to grid A, the dot-dashed line to grid C, and the dotted line to grid D.}
 \label{fig:cut_comp_backward}
\end{figure}

To see the effect of the grid on the conversion we plot a selection of forward converted spectra using all four grids divided by the corresponding emulated spectra in Fig.~\ref{fig:cut_comp_forward} and the backward converted spectra divided by the corresponding linear spectra in Fig.~\ref{fig:cut_comp_backward}. In grid D the power spectrum is only filled up to $k=1.4\,h\,\mathrm{Mpc}^{-1}$ before the conversion, higher modes are set to zero.
In the backward conversion the power spectra start to differ around $k=0.9\,h\,\mathrm{Mpc}^{-1}$, where the result from grid D starts to overshoot the others. In the forward conversion the differences are much more prominent. For redshift 0 the spectrum from grid C exhibits a clear increase in broad-band power on all scales, because at $z=0$ the region with $1.4\,h\,\mathrm{Mpc}^{-1}<k\leq3.4\,h\,\mathrm{Mpc}^{-1}$ contains more power than the region with $k\leq1.4\,h\,\mathrm{Mpc}^{-1}$. This additional power couples to all scales in the forward conversion. However, for redshifts greater than $1$ the spectra appear to be consistent up to $k=0.8\,h\,\mathrm{Mpc}^{-1}$ and grids A, B, and D appear to be consistent at all redshifts. Table~\ref{table:grid-log-distances} lists the log-distances up to $k=1.0\,h\,\mathrm{Mpc}^{-1}$ for grids A, C, and D with respect to grid B. One can see, that the log-distances are comparable with the log-distances of grid B with respect to the emulated and linear spectra listed in Table~\ref{table:log-distances} in Appendix~\ref{sec:log-distances}. We therefore conclude that the calculation is consistent between the investigated grids within the precision of our model.
There is however a systematic trend apparent in Fig.~\ref{fig:cut_comp_forward} that an increase in total power due to a greater covered $k$-range leads to an increase in broad-band power in the forward converted spectra. We therefore expect the validity of our model to break down for higher high dynamic ranges as the total power of the Cosmic matter spectrum diverges, since its spectral index towards high $k$ is above $-3$. The problem with a spectral index above $-3$ can be seen best in the angle integrated conversion formula Eq.~\eqref{eq:angles_integrated}.

\begin{table}
\caption{Log-distances for $k\leq1.0\,h\,\mathrm{Mpc}^{-1}$ between the converted spectra at different grids. The distances are with respect to grid B (the grid used in Sect.~\ref{sec:matter_spectrum}); ``(fwd)'' indicates forward conversion, ``(bwd)'' indicates backward conversion.}              
\label{table:grid-log-distances}      
\centering                                      
\begin{tabular}{c  c  c  c  c  c  c }          
\hline\hline
 redshift & 0 & 0.5 & 1.0 & 1.5 & 2 & 3\\    
\hline                                   
 grid A (fwd) & 0.4 & 0.16 & 0.07 & 0.04 & 0.05 & 0.04 \\      
 grid C (fwd) & 1.9 & 0.7 & 0.3 & 0.12 & 0.07 & 0.03\\
 grid D (fwd) & 0.18 & 0.2 & 0.19 & 0.15 & 0.12 & 0.08\\
 grid A (bwd) & 0.8 & 0.18 & 0.12 & 0.09 & 0.06 & 0.03 \\      
 grid C (bwd) & 0.3 & 0.2 & 0.2 & 0.10 & 0.07 & 0.04\\
 grid D (bwd) & 0.7 & 0.4 & 0.2 & 0.18 & 0.13 & 0.07\\
\hline                                             
\end{tabular}
\end{table}

\subsection{Strength of the smearing of BAO}
\label{sec:smear_strength}

We estimate the strength of the smearing of baryon acoustic oscillations by comparing to a smoothing (with respect to $\log(k)$) of the linear spectrum using a Gaussian kernel,
\begin{equation}
 g(\log(k);\sigma_g) = \frac{1}{\sqrt{2\pi\sigma_g^2}} \exp\left(-\frac{1}{2} \left(\frac{\log(k)}{\sigma_g}\right)^2\right),
\end{equation}
where $\sigma_g$ is the smoothing length in terms of $e$-folds. $\sigma_g$ is a dimensionless distance in $\log(k)$.
The smoothed spectrum is therefore
\begin{equation}
 P_\mathrm{sm}(k;\sigma_g) = \int\!\!\mathrm{d}\log(q)\ g(\log(k/q);\sigma_g)\,P_\mathrm{lin}(q).
\end{equation}
We can compare $P_\mathrm{nl}/P_\mathrm{nl,nw}$ with $P_\mathrm{sm}(\sigma_g)/P_\mathrm{lin,nw}$ in order to find a $\sigma_g$ where the smoothing appears to be similar. The higher $\sigma_g$, the stronger the smoothing.
Here, $P_\mathrm{nl}$ denotes the non-linear power spectrum with BAO, $P_\mathrm{nl,nw}$ the non-linear power spectrum without BAO, and $P_\mathrm{lin,nw}$ the linear power spectrum without BAO. The best value for $\sigma_g$ at each redshift is found by visual comparison in the absence of a more rigorous criterion.
We list the best fit values for different redshifts for the 3PT spectra as well as the forward converted spectra in Table~\ref{table:smoothing}. The values for the CosmicEmu spectra are missing because the CosmicEmu code did not allow us to calculate non-linear power spectra without BAO.

\begin{table}
\caption{Best fit smoothing scales of a Gaussian smoothing on log-scale that emulate the smearing of BAO for the 3PT spectra and the forward converted spectra a different redshifts. A higher smoothing scale means stronger smoothing. The 3PT result is smoothed stronger than the forward-converted spectrum at all redshifts.}              
\label{table:smoothing}      
\centering                                      
\begin{tabular}{c  c  c }          
\hline\hline
   redshift & 3PT smoothing & forward conversion \\    
   & scale & smoothing scale \\ 
\hline                                   
 7 & 0.03 & <0.01\\
 6 & 0.03 & <0.01\\
 5 & 0.035 & <0.01\\
 4 & 0.035 & <0.01\\
 3 & 0.045 & 0.015\\
 2 & 0.055 & 0.015\\
 1.5 & 0.07 & 0.02\\
 1.2 & 0.07 & 0.02\\
 1.0 & 0.075 & 0.025\\
\hline
\end{tabular}
\end{table}

\subsection{Large-scale bias}
\label{sec:bias-factor}

\cite{Neyrinck-2009} find a bias factor between the the power spectra of the density contrast and the logarithmic density, which they fit to $\exp\!\left( -\mathrm{Var}[\log(1+\delta)]\right)$ with high accuracy up to $z\approx1.2$. If the density field follows log-normal statistics the correlation functions of the density and log-density are related by
\begin{equation}
\begin{split}
 \left\langle \rho(\vec{x}) \rho(\vec{y}) \right\rangle_{\mathcal{P}(\rho)} =  & \exp\!\left( \frac{1}{2} \left\langle s(\vec{x})^2 \right\rangle_{\mathcal{P}(s)} + \frac{1}{2} \left\langle s(\vec{y})^2 \right\rangle_{\mathcal{P}(s)}\right) \times\\
 & \exp\!\left(\left\langle s(\vec{x})\,s(\vec{y}) \right\rangle_{\mathcal{P}(s)} \right),
 \end{split}
\end{equation}
which simplifies under the assumption of statistical homogeneity and a zero mean for $s$ to
\begin{equation}
 \begin{split}
  \left\langle \rho(\vec{x}) \rho(\vec{y}) \right\rangle_{\mathcal{P}(\rho)} =  & \exp\!\left( \mathrm{Var}[s]\right) \times\\
 & \exp\!\left(\left\langle s(\vec{x})\,s(\vec{y}) \right\rangle_{\mathcal{P}(s)} \right),
 \label{eq:bias_factor}
\end{split}
\end{equation}
where $\mathrm{Var}[s] = \left\langle s(\vec{x})^2 \right\rangle_{\mathcal{P}(s)}$. This prefactor is the square of the expectation value of $\rho$,
\begin{equation}
 \left\langle \rho(x) \right\rangle_{\mathcal{P}(\rho)} = \exp\!\left( \frac{1}{2} \left\langle s(\vec{x})^2 \right\rangle_{\mathcal{P}(s)} \right) = \sqrt{\exp\!\left( \mathrm{Var}[s]\right)}.
\end{equation}
In Sect.~\ref{sec:matter_spectrum} of this work, we compared the spectra of the $\delta$ and $e^s/\left\langle e^s\right\rangle$ since the density contrast is defined around a mean of $1$. In the backward conversion, the mean of the resulting log-density field was not zero. Eq.~\ref{eq:backward_mono} determined it as
\begin{equation}
 m = -\frac{1}{2}\left\langle s(\vec{x})^2 \right\rangle_{\mathcal{P}(s)}
\end{equation}
for all redshifts (to 0.1\% precision). However, this prefactor should not appear in the work of \cite{Neyrinck-2009}, since they compare the power spectra of $\delta$ and $s = \log(1+\delta)$, where $\delta = \rho/\rho_0 - 1$. This means that the prefactor in Eq.~\eqref{eq:bias_factor} is already divided out.

There is an additional bias apart from this factor, which can be seen best in the low redshift panels in the left part of Fig.~\ref{fig:high_over_all}. Here the forward converted power spectrum exceeds the original power spectrum even at the lowest $k$-bin. This is not due to the prefactor presented in the previous paragraph, but simply to mode coupling in the forward conversion. The inverse of this effect is present in the backward conversion, where the backward (i.e.,~log-transformed) power spectrum undershoots the linear power spectrum at the lowest $k$-bin. This factor is, however, much weaker than what \cite{Neyrinck-2009} find. For completeness, we list the factors between the lowest $k$-bins, i.e.,~$P_\mathrm{backward}(k\rightarrow 0)/P_\mathrm{emulated}(k\rightarrow 0)$, in Table~\ref{table:low_k_factors}.
In conclusion, the log-normal model offers no insight to the bias factor found by \cite{Neyrinck-2009}.

\begin{table}
\caption{Large-scale bias factors due to mode coupling. Listed are the values of the backward converted power spectra for the lowest $k$-bin divided by the power in the lowest $k$-bin in the respective emulated spectrum.}              
\label{table:low_k_factors}      
\centering                                      
\begin{tabular}{c c c }          
\hline\hline
   redshift & $P_\mathrm{backward}(k\rightarrow 0)/$ & $\mathrm{Var}[\log(1+\delta)]$ \\    
   & $P_\mathrm{emulated}(k\rightarrow 0)$ & \\ 
\hline                                   
 7 & 1.00 & 0.095\\
 6 & 1.00 & 0.124\\
 5 & 1.00 & 0.168\\
 4 & 0.99 & 0.24\\
 3 & 0.99 & 0.38\\
 2 & 0.98 & 0.65\\
 1.5 & 0.97 & 0.89\\
 1.2 & 0.97 & 1.10\\
 1.0 & 0.96 & 1.27\\
 0.9 & 0.96 & 1.34\\
 0.8 & 0.95 & 1.44\\
 0.7 & 0.95 & 1.55\\
 0.6 & 0.95 & 1.66\\
 0.5 & 0.94 & 1.78\\
 0.4 & 0.94 & 1.90\\
 0.3 & 0.93 & 2.04\\
 0.2 & 0.93 & 2.17\\
 0.1 & 0.92 & 2.32\\
 0.0 & 0.91 & 2.47\\
\hline
\end{tabular}
\end{table}

\subsection{Generating matter densities in spaces spanning in redshift}
\label{sec:generate_matter}

In Sect.~\ref{sec:matter_spectrum} we have established that the power spectra of the exponentiated linear density contrast agree with the emulated spectra to a reasonable accuracy down to redshift 1 and $k\leq1.0\,h\,\mathrm{Mpc}^{-1}$.
Therefore, the log-density at different redshifts is (in a statistical average) related by a simple global prefactor. This enables us to formulate a local function which translates the density between different redshifts to better accuracy than linear theory. We let $G(z;z_0)$ be the growth factor between redshift $z_0$ and redshift $z$ and $\delta(z)$ the density contrast at a given redshift. Then we have
\begin{equation}
\begin{split}
1 + \delta(z) \approx & \exp\!\left\{-\frac{1}{2}G(z;z_0)^2\, \mathrm{Var}\!\left[\log(1+\delta(z_0))\right] \right\}\times \\
                   & \exp\!\left\{G(z;z_0)\, \log(1+\delta(z_0))\right\},
\end{split}
\label{eq:density_trafo}
\end{equation}
where $\mathrm{Var}[\cdot]$ is the variance in one cell.

Using this formula one can easily generate a lognormal field that behaves like the matter density contrast to a good accuracy. One simply takes $z_0$ to be sufficiently high so that $\mathrm{Var}\!\left[\delta(z_0)\right]\ll 1$ and $\log(1+\delta(z_0)) \approx \delta(z_0)$. At such a redshift a Gaussian random field generated from the matter power spectrum desribes the statistics of $\delta(z_0)$ very well. By applying Eq.~\eqref{eq:density_trafo} using a position dependent redshift $z(\vec{x})$,
\begin{equation}
 \begin{split}
\delta(\vec{x}) \approx & \exp\!\left\{-\frac{1}{2}G(z(\vec{x});z_0)^2\, \mathrm{Var}\!\left[\delta(z_0)\right] \right\}\times \\
                   & \exp\!\left\{G(z(\vec{x});z_0)\, \delta(z_0)\right\} - 1,
\end{split}
\end{equation}
one can now generate a log-normal matter density contrast that follows the appropriate non-linear matter power spectrum in each position (or redshift slice).

\end{appendix}

\end{document}